\documentclass[final,3p,times]{elsarticle}

\usepackage{graphicx,epsfig}
\usepackage{natbib}
\usepackage{amssymb}

\newcommand{\beq}{\begin{equation}} 
\newcommand{\eeq}{\end{equation}} 
\newcommand{\beqa}{\begin{eqnarray}} 
\newcommand{\eeqa}{\end{eqnarray}} 
\newcommand{\bea}{\begin{array}} 
\newcommand{\ea}{\end{array}} 

\newcommand{\dd}{{\rm d}}
\newcommand{\pl}{\partial}
\newcommand{\inta}{\int_{-i\infty}^{+i\infty}} 
\newcommand{\lag}{\langle} 
\newcommand{\rag}{\rangle}
\newcommand{\law}{\stackrel{\rm law}{=}}
\newcommand{\ii}{{\rm i}}

\newcommand{\cP}{{\cal P}}

\newcommand{\tG}{\tilde{G}}
\newcommand{\Hi}{H_{\infty}}

\newcommand{\tD}{\tilde{\Delta}}

\newcommand{\gam}{\gamma}

\newcommand{\bp}{{\overline p}}

\newcommand{\tE}{\tilde{E}}

\newcommand{\Ai}{\mbox{Ai}}

\journal{Physica A}

\begin{document}

\begin{frontmatter}

\title{Ballistic aggregation for one-sided Brownian initial velocity}

\author{Patrick Valageas}

\address{Institut de Physique Th{\'e}orique, CEA Saclay, 
91191 Gif-sur-Yvette, France}

\begin{abstract}
We study the one-dimensional ballistic aggregation process in the continuum
limit for one-sided Brownian initial velocity (i.e. particles merge when they 
collide and move freely between collisions, and in the continuum limit the initial 
velocity on the right side is a Brownian motion that starts from the origin $x=0$). 
We consider the cases where the left side is either at rest or empty at $t=0$. 
We derive explicit expressions for the velocity distribution and the mean density 
and current profiles built by this out-of-equilibrium system. We find that on the 
right side the mean density remains constant whereas the mean current is uniform 
and grows linearly with time. 
All quantities show an exponential decay on the far left. We also obtain the 
properties of the leftmost cluster that travels towards the left. 
We find that in both cases relevant lengths and masses scale as $t^2$ and the 
evolution is self-similar.
\end{abstract}

\begin{keyword}
Adhesive dynamics \sep Ballistic aggregation \sep Inviscid Burgers equation \sep 
non-equilibrium statistical mechanics


% MSC codes here, in the form: \MSC code \sep code
% or \MSC[2008] code \sep code (2000 is the default)

\end{keyword}

\end{frontmatter}

\section{Introduction}
\label{sec:intro}

We consider in this article the continuum limit of a one-dimensional ballistic 
aggregation process, for the case of Brownian initial velocities (i.e. the initial
velocity field is a Brownian motion). In such a model, 
point particles of identical mass $m$ move on a line and perform completely inelastic 
collisions, that is, in binary collisions particles (or clumps) merge to form a 
single larger aggregate under conservation of mass and momentum (but dissipation of 
energy). Between collisions clumps move at constant velocity (free motion). 
Thus, without external forcing, the stochasticity is only due to the randomness of
the initial velocities. This model was introduced in \cite{Carnevale1990}, for the
case of uncorrelated initial velocities (i.e. white-noise case in the continuum limit),
as a simple test-case for scaling arguments used in more general hydrodynamical
or statistical systems. Indeed, this ballistic aggregation process can be seen
as a simple model for the merger of coherent structures, such as vortices, thermal
plumes, or cosmic dust into planetesimals within proto-planetary disks.

In this context it is natural to investigate the late-time asymptotic scaling regime
obtained for the case of uncorrelated initial velocities.
Thus, one finds that the average cluster mass grows with time as $t^{2/3}$
with a large-mass tail for the universal mass distribution of the form $e^{-m^3/t^2}$
\cite{Carnevale1990,Frachebourg1999,Frachebourg2000a}. When the number of particles
is finite, at long times the system reaches a stationary ``fan'' state, where the
velocities of the final clusters increase from left to right. This final state
also shows many universal properties, such as the number and size distributions
of final clusters and the size of the leftmost and rightmost clusters
\cite{Suidan2000,Majumdar2008}. On the other hand, when the initial particle 
velocities are
given by a Brownian motion this ballistic aggregation process can be related to
a simple additive coalescent model (which does not take into account positions nor
velocities), where each pair of clusters merges with a rate proportional to its total 
mass, independently of other pairs \cite{Bertoin2000}. This also provides results
for the statistics of dislocation of clusters in the time-reversed fragmentation
process. 

In the continuum limit, $m\rightarrow 0$ at fixed uniform initial density $\rho_0$, it 
is well known that this system can be mapped onto the Burgers equation in the 
inviscid limit \cite{Burgersbook,Gurbatov1991,Kida1979,Frachebourg2000a},
\beq
\frac{\pl v}{\pl t} + v \frac{\pl v}{\pl x} = \epsilon \frac{\pl^2 v}{\pl x^2}
\hspace{1cm} \mbox{with} \hspace{1cm} \epsilon \rightarrow 0^+ .
\label{Burg}
\eeq
Then, shock locations in the Eulerian velocity field $v(x,t)$ describe particle 
aggregates of finite mass whereas regular points correspond to infinitesimal 
particles. It is clear that away from shocks Eq.(\ref{Burg}) corresponds to free 
motion (in the limit $\epsilon \rightarrow 0^+$ where the right-hand side vanishes) 
and it can be shown that shocks conserve momentum \cite{Burgersbook}, which explains 
the relation with the ballistic aggregation process.

The Burgers equation (\ref{Burg}) itself is a nonlinear evolution 
equation that appears in many physical problems, such as turbulence studies 
\cite{Burgersbook,Kida1979}, the propagation of nonlinear acoustic
waves \cite{Gurbatov1991}, or the formation of large-scale structures in cosmology
\cite{Gurbatov1989,Vergassola1994}, see the recent review \cite{Bec2007} for
a detailed discussion. In particular, the study of the statistical properties of
the dynamics, starting with random Gaussian initial conditions, is also referred to
as ``decaying Burgers turbulence'' in the hydrodynamical context \cite{Gurbatov1997},
or as the ``adhesion model'' in the cosmological context \cite{Gurbatov1989}.
In these frameworks, where the random initial velocity applies over all space, and 
may be homogeneous or only have homogeneous increments, one is interested in 
Eulerian quantities such as the velocity structure functions, the $n$-point velocity 
distributions, the matter density distribution, the mass function of shocks,
or Lagrangian quantities such as the distribution of the displacement field.
Then, it is customary to consider power-law initial energy spectra, 
$E_0(k) \propto k^n$,
where the initial velocity fluctuations scale as $x^{-(n+1)/2}$ over size $x$
(the white-noise case is $n=0$ and the Brownian case is $n=-2$). Then, for $-3<n<1$,
a self-similar evolution develops \cite{Gurbatov1991,Molchanov1995}. The integral 
scale of turbulence, which measures the typical distance between shocks and the 
correlation length, grows as $L(t) \sim t^{2/(n+3)}$ whereas the tails of the 
cumulative shock distribution and velocity distribution satisfy 
$\ln[n(>m)] \sim -m^{n+3}$, $\ln[n(>|v|)] \sim -|v|^{n+3}$, for $m\rightarrow
\infty,|v|\rightarrow\infty$, see \cite{She1992,Molchan1997}.
In such a context, the white-noise case, $n=0$, corresponds to initial velocity 
fluctuations that are dominated by high wavenumbers, whereas they are governed by
low wavenumbers in the Brownian case.
Then, this latter case is of particular interest since in many hydrodynamical
systems the power is generated by the larger scales. For instance, the Kolmogorov
spectrum of turbulence, $E(k)\propto k^{-5/3}$, shows such an infrared divergence,
whereas in cosmology the velocity fluctuations are also governed by scales that are
larger than the nonlinear scales where structures have already formed in the density
field. 

The connection with the Burgers equation (\ref{Burg}) allows us to derive many results
for the continuum limit of the ballistic aggregation process, taking advantage of 
its well-known Hopf-Cole solution \cite{Hopf1950,Cole1951}. (This also corresponds
to the late-time evolution of the system if we keep a finite particle mass.)
In particular, using the geometrical description of this solution in terms of 
first-contact points between the initial velocity potential and parabolas
\cite{Burgersbook}, as recalled below in (\ref{paraboladef}), 
or the equivalent description in terms of the
convex hull of the Lagrangian potential \cite{Bec2007}, it is possible to derive 
closed analytical results for the specific cases of white-noise and Brownian initial 
velocities. Indeed, in these two cases the velocity or potential fields are Markovian
which allows us to greatly simplify the analysis 
\cite{Bertoin1998,Frachebourg2000,Valageas2008}.
Moreover, a specific property obtained for Brownian initial conditions is that the 
inverse Lagrangian map, $x\mapsto q(x,t)$, where $q$ is the initial Lagrangian 
position of the particle located at $x$ at time $t$, keeps homogeneous increments
at all times (on the right side for one-sided initial conditions and far from the 
origin for two-sided initial conditions) \cite{Bertoin1998,Valageas2008}.
This can actually be extended to some L{\'e}vy processes with no positive jumps
\cite{Bertoin1998}.

Then, for both white-noise and Brownian one-sided cases, \cite{Bertoin2001}
found that the flux through the origin is a pure jump time-inhomogeneous Markov 
process and obtained its statistical distribution. 
While for the white-noise case clusters (shocks) are in finite number per unit
length, which implies that the mass that has flowed to the the left side increases
through a finite number of jumps per unit time, for the Brownian case clusters are
dense (on the right side), which implies that on any time interval some small-mass 
clusters have crossed the origin \cite{She1992,Bertoin2001}. 
Some properties of the limit clusters travelling
towards the left were obtained in \cite{Isozaki2006} for the white-noise case,
while for the Brownian case it was found that shocks are dense to the right
of the leftmost cluster, as on the right side, but no results for the statistics
of the latter were obtained. 
Statistical properties of limit clusters (i.e. at the infinite time limit) were 
also obtained in \cite{Winkel2002} for initial velocities that are given by a 
L{\'e}vy process with no positive jumps.

For the white-noise case, \cite{Frachebourg2000a,Frachebourg2001} studied the
late-time dynamics reached when the ``excited'' particles are restricted to the 
semi-infinite right side, or to a finite interval, and expand into empty space 
or a medium at rest. Many explicit analytical results can be derived in the continuum 
limit \cite{Frachebourg2001}. 
If the initial ``excited'' interval is finite and surrounded by empty space, 
the late-time evolution is ballistic and the characteristic length scales as 
$L(t) \propto t$. Indeed, since the total mass is finite, the system eventually 
reaches a ``fan'' state with a finite number of clusters that move freely without 
anymore collisions. If the ``excited'' particles expand into a medium of uniform
density at rest, the latter slows down the motion and the characteristic length 
scales as $L(t) \propto t^{1/2}$. For one-sided initial conditions, where the
initial white-noise velocities apply to the semi-infinite right axis, one recovers
the scaling law associated with the homogeneous turbulent case recalled above,
$L(t) \propto t^{2/3}$ and $M(t)\propto t^{2/3}$. Thus, nontrivial mass density 
and current profiles develop over this scale, with a power-law tail 
$\rho\sim |x|^{-3}$ on the far left. In addition, other nontrivial asymptotic mass 
profiles are obtained over scales $L_{\alpha}(t) \propto t^{\alpha}$, with
$2/3<\alpha\leq 1$, that interpolate between the natural scaling $\propto t^{2/3}$ 
and the ballistic regime $\propto t$. This corresponds to forerunners that carry
a mass $\propto t^{2(1-\alpha)}$, interpolating between masses of order $t^{2/3}$ and 
of order unity. Finally, if the ``excited'' particles expand into a filled medium 
at rest, other profiles develop on the natural scale $L(t) \propto t^{2/3}$,
but there is no more propagation on larger scales, $\alpha>2/3$, because of the
slowing down by the left-side particles, and the density decays exponentially fast
on the left side.

In this article, we study how these results are modified when the initial
velocities on the right side are given by a Brownian motion, instead of a white-noise
spectrum, and we also consider the statistical properties of the leftmost 
(``leader'') cluster that has formed on the left side. As noticed above, the Brownian
case is a template for large-scale forcing as opposed to small-scale forcing in the
initial velocity field. Thus, we consider the case where the initial 
velocity field, $v_0(q)$ at time $t=0$, is a Brownian motion on the semi-infinite 
axis $q\geq 0$, while the initial density $\rho_0$ is constant over $q\geq 0$,
and we write
\beq
\rho(q,0)=\rho_0, \;\;\; v_0(q)=  \int_0^q\dd q' \, \xi(q') \;\;\; \mbox{and} \;\;\; 
\psi_0(q)= \int_0^q\dd q' \int_0^{q'}\dd q'' \, \xi(q'') \;\;\;\;\; \mbox{over} 
\;\; q\geq 0  .
\label{def} \\
\eeq
Here we introduced the velocity potential $\psi(x,t)$, with $v = \pl\psi/\pl x$,
and a Gaussian white-noise $\xi(q)$, which we normalize by
\beq
\lag\xi(q)\rag = 0, \;\;\;\; \lag\xi(q)\xi(q')\rag = D \, \delta(q-q') ,
\;\;\; \mbox{whence} \;\;\; \lag v_0(q)^2\rag = D q \;\; 
\mbox{in the Brownian region} , 
\label{Ddef}
\eeq
where $\lag .. \rag$ is the average over all realizations of $\xi$. 
In Eq.(\ref{def}) we normalized the initial velocity and potential
by $v_0(0)=0$ and $\psi_0(0)=0$ at the origin.
We consider two cases, ``$F$'' and ``$E$'', where the left semi-infinite axis,
$q<0$, is either filled with particles at the same density $\rho_0$ but with zero 
initial velocity (medium at rest), or empty (zero density, $\rho=0$). Therefore, 
we complete the definition (\ref{def}) by
\beqa
\mbox{case } ``F'' & : & \rho(q,0)=\rho_0, \;\; v_0(q)=0 \;\; \mbox{and} \;\; 
\psi_0(q)=0 \;\; \mbox{over} \;\; q < 0 , \label{Fdef} \\
\mbox{case } ``E'' & : & \rho(q,0)=\rho_0, \;\; v_0(q)=v_- \;\; \mbox{and} \;\; 
\psi_0(q)=v_- q \;\; \mbox{over} \;\; q < 0 , \;\; \mbox{with} \;\;  
v_-\rightarrow -\infty . \label{Edef}
\eeqa
Here, as in \cite{Isozaki2006}, we used the fact that the empty case, ``$E$'', 
of (\ref{Edef}), can be obtained by keeping the same uniform initial density 
$\rho_0$ over $q<0$, while giving to these particles a velocity $v_-$ that goes to 
$-\infty$. Then, these particles immediately escape to the infinite left at $t=0^+$ 
and the Brownian particles with $q\geq 0$ spread into empty space.
These initial conditions can be summarized by stating that the 
initial velocity potential $\psi_0$ is either continuous and constant out of the 
Brownian region (filled case ``$F$'') or goes to $+\infty$ (empty case ``$E$''). 

At any point $x<0$ on the left part, the 
system remains unchanged (at rest or empty) until some particles that originate from 
the right-side Brownian region have managed to travel down to position $x$. 
Note that once some particles have entered the left part they will keep travelling 
with a negative velocity forever. However, their velocity can change as they may 
overtake slower particles or may be overtaken by faster particles that escaped at a 
later time from the Brownian region. Since particles do not cross, the leftmost 
cluster is associated with the particle, $q=0$, that was initially at 
the left boundary of the Brownian domain. Once the latter has entered the left side, 
it keeps moving to the left and in case $F$ it draws along all the matter that was 
initially at rest.

In the context of the ballistic aggregation process studied in this article,
and contrary to the hydrodynamical context where the Burgers equation (\ref{Burg})
is used to investigate statistically homogeneous turbulence,
the systems defined by Eqs.(\ref{def})-(\ref{Edef}) are clearly 
statistically inhomogeneous and a current develops towards the left side as 
particles escape into the left part of the system and then keep travelling to the 
left forever. Therefore, we mainly focus on quantities that express this
out-of-equilibrium propagation of matter towards the left. 
Indeed, the conditional probabilities to the right of the leftmost cluster are
identical to the ones obtained for the two-sided Brownian-motion initial 
velocity \cite{Valageas2008}. In particular, the distribution of velocity increments
and of the matter density are the same on the right side $x>0$, see also
\cite{Bertoin1998}.

We recall in section~\ref{sec:kernels} the geometrical construction in terms
of first-contact parabolas of the solution of Eq.(\ref{Burg}) and the associated
Brownian propagators. Next, we first consider the case ``$F$'' of a filled left-side
at rest, and we study the velocity distribution $p_x(v)$ as well as the probability
$p_x^{\rm shocked}$ that matter from the right side has already reached the position
$x$ on the left side by time $t$. We consider the mean density profile and current
in section~\ref{subsec:density-profile}. Then, we derive the Lagrangian displacement
field in section~\ref{subsec:Lagrangian_displacement} and we obtain in
section~\ref{subsec:Leftmost_cluster} the properties of the leftmost cluster.
Finally, we consider the case ``$E$'' of the empty left side in 
section~\ref{sec:case/E}.

\section{Geometrical construction and Brownian propagators with parabolic absorbing 
barrier}
\label{sec:kernels}

As recalled above, in the continuum limit, $m\rightarrow 0$ at fixed density $\rho_0$,
the ballistic aggregation dynamics is fully described by the Burgers equation 
(\ref{Burg}) in the limit of zero viscosity. As is well known 
\cite{Hopf1950,Cole1951}, substituting for the velocity potential 
$\psi(x,t)$, with $v = \pl\psi/\pl x$, and making the change of variable 
$\psi(x,t)=-2\nu\ln\theta(x,t)$, transforms the nonlinear Burgers equation into the 
linear heat equation. This provides the explicit solution of Eq.(\ref{Burg}) for
any initial condition, and in the limit $\epsilon \rightarrow 0^+$ a saddle-point 
method gives
\beq
\psi(x,t) = \min_q \left[ \psi_0(q) + \frac{(x-q)^2}{2t} \right]
\hspace{0.5cm} \mbox{and} \hspace{0.5cm} v(x,t) = \frac{x-q(x,t)}{t} ,
\label{psinu}
\eeq
where we introduced the Lagrangian coordinate $q(x,t)$ defined by
\beq
\psi_0(q) + \frac{(x-q)^2}{2t} \hspace{0.5cm} \mbox{is minimum at the point} 
\hspace{0.5cm} q = q(x,t) .
\label{qmin}
\eeq
The Eulerian locations $x$ where there are two solutions $q_-<q_+$ to the 
minimization problem (\ref{qmin}) correspond to shocks (and all the matter initially 
between $q_-$ and $q_+$ is gathered at $x$), that is to clumps of particles of 
finite mass. The application $q \mapsto x(q,t)$ is usually called the Lagrangian 
map, and $x \mapsto q(x,t)$ the inverse Lagrangian map (which is discontinuous at
shock locations). For the case of Brownian initial velocity that we consider
in this paper, it is known that the set of regular Lagrangian points has
a Hausdorff dimension of $1/2$ \cite{Sinai1992}, whereas shock locations
are dense in Eulerian space \cite{Sinai1992,She1992}, in the Brownian region.

As is well known \cite{Burgersbook}, the minimization problem (\ref{qmin})
has a nice geometrical solution. Indeed, let us consider the 
downward\footnote{In the literature one often defines the velocity potential
as $v=-\pl_x\psi$, which leads to upward parabolas. Here we prefer to define
$v=\pl_x\psi$ to follow \cite{Valageas2008}.} parabola $\cP_{x,c}(q)$ centered at 
$x$ and of maximum $c$, i.e. of vertex $(x,c)$, of equation
\beq
\cP_{x,c}(q) = - \frac{(q-x)^2}{2 t} + c .
\label{paraboladef}
\eeq
Then, starting from below with a large negative value of $c$, such that the
parabola is everywhere well below $\psi_0(q)$ (this is possible thanks to the
scaling $\psi_0(\lambda q) \law \lambda^{3/2} \psi_0(q)$ for the integral $\psi_0$ 
of the Brownian motion, which shows that $|\psi_0(q)|$ only grows as $q^{3/2}$
at large $q$), we increase $c$ until the two curves touch one another.
Then, the abscissa of the point of contact is the Lagrangian coordinate
$q(x,t)$ and the potential is given by $\psi(x,t)=c$.

This geometrical construction clearly shows that a key quantity is the conditional 
probability density, 
\newline $K_{x,c}(q_1,\psi_1,v_1;q_2,\psi_2,v_2)$, 
for the Markov process 
$q\mapsto\{\psi_0(q),v_0(q)\}$, starting from $\{\psi_1,v_1\}$ at $q_1 \geq 0$, 
to end at $\{\psi_2,v_2\}$ at $q_2 \geq q_1 \geq 0$, while staying above the 
parabolic barrier, $\psi_0(q)>\cP_{x,c}(q)$, for $q_1\leq q\leq q_2$. Following 
\cite{Frachebourg2000} (who studied the case of two-sided white-noise initial 
velocity) and \cite{Valageas2008} (who studied the case of two-sided Brownian initial 
velocity), we shall obtain the properties of the system from this propagator.
It was derived in \cite{Valageas2008} who obtained
\beq
K_{x,c}(q_1,\psi_1,v_1;q_2,\psi_2,v_2) \, \dd \psi_2 \dd v_2 = 
e^{-\tau/\gamma^2+(u_2-u_1)/\gamma} \, G(\tau;r_1,u_1;r_2,u_2) \, \dd r_2 \dd u_2 ,
\label{KG1}
\eeq
with
\beq
\tau=\gamma^2 (Q_2-Q_1) , \;\;\; 
r_i=2\gamma^3\left[ \Psi_i+\frac{(Q_i-X)^2}{2}-C \right], 
\;\;\; u_i= 2\gamma (V_i+Q_i-X) .
\label{riui}
\eeq
Here we introduced the dimensionless coordinates (which we shall note by capital 
letters in this article)
\beq
Q=\frac{q}{\gam^2} , \;\; X= \frac{x}{\gam^2} , \;\; V=\frac{tv}{\gam^2} , \;\; 
\Psi = \frac{t\psi}{\gamma^4} , \;\;  C = \frac{t c}{\gamma^4} , \;\; \mbox{with} 
\;\; \gam = \sqrt{2D} \, t , \;\; \mbox{whence} \;\;\; X=Q+V  \;\;\;
\mbox{for regular points} .
\label{QXdef}
\eeq
For completeness, we give in Appendix~\ref{app:Brownian_propagators} the expression 
of the reduced propagator $G$ and of two associated kernels.

We can note that, thanks to the scale invariance of the
Brownian motion, the scaled initial potential $\psi_0(\lambda q)$ has the same
probability distribution as $\lambda^{3/2} \psi_0(q)$, for any $\lambda>0$.
Then, using the explicit solution (\ref{psinu}) we obtain the scaling laws
\beq
\psi(x,t) \law t^3 \psi(x/t^2,1) , \;\;\; v(x,t) \law t v(x/t^2,1) , \;\;\;
q(x,t) \law t^2 q(x/t^2,1) ,
\label{scalings}
\eeq
where $\law$ means that both sides have the same probability distribution.
Indeed, the initial conditions on the left side, $\psi_0(q)=0$ or 
$\psi_0(q)=+\infty$, do not break these scalings. This implies that all 
our results can be written in terms of the dimensionless variables (\ref{QXdef}), 
as we shall check below. This would no longer hold if the Brownian domain were 
restricted to a finite interval $[-L,L]$, since the size $L$ would add a new length 
scale into the problem which can give rise to other scalings. In particular, at late
times one would find a simple ballistic propagation $|x| \propto t$ to the left
or right side in the empty case, when all high-velocity particles have already 
escaped from the Brownian domain.

Finally, we may note that we defined the continuum limit as $m\rightarrow 0$ at
fixed density $\rho_0$. The same limit also describes the large time behavior
of the system (i.e. $t\rightarrow \infty$) at fixed reduced lengths and masses
$X=x/\gam^2$ and $M=m/(\rho_0\gam^2)$, as in (\ref{QXdef}). 
In the first point of view, we consider the properties of the system at any finite 
time, arising from a distribution of infinitesimal particles, whereas in the second
point of view we keep the discrete nature of the initial system but we look at the 
asymptotic late-time distribution over large scales and masses that grow as $t^2$ 
(so that discrete effects become subdominant); see \cite{Frachebourg2000a} for
more detailed analysis in the case of white-noise initial velocity.

\section{Case ``$F$'': expansion into an uniform medium at rest}
\label{sec:case/F}

We first investigate the case ``$F$'', defined in (\ref{Fdef}), where the 
left side, $q<0$, is initially filled with particles at the same uniform density 
$\rho_0$ with zero velocity.

\subsection{Eulerian velocity distribution $p_x(v)$ and probability 
of having been shocked $p_x^{\rm shocked}$}
\label{subsec:Eulerian_velocity_distribution}

\begin{figure}
\begin{center}
\epsfxsize=9.5 cm \epsfysize=5.8 cm {\epsfbox{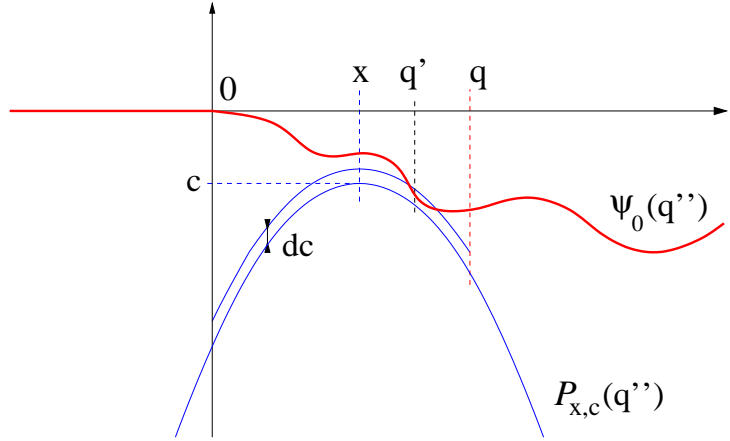}}
\end{center}
\caption{Geometrical interpretation of the initial conditions $\psi_0(q'')$ 
associated with the probability $p_x(0\leq q'\leq q,c)\dd c$. The Brownian curve
$\psi_0(q'')$ is everywhere above the parabola $\cP_{x,c}$ and goes below 
$\cP_{x,c+\dd c}$ somewhere in the range $0\leq q''\leq q$. On the left side,
$\psi_0(q'')=0$ and it is also above the parabola $\cP_{x,c}$.
To obtain the cumulative probability, $p_x(0\leq q'\leq q)$, we must then integrate 
over the height $c$ of the parabola.}
\label{figP1}
\end{figure}

We first consider the one-point velocity distribution, $p_x(v)$,
at the Eulerian location $x$, as well as the distribution, $p_x(q)$, of the 
Lagrangian coordinate $q(x,t)$ associated with the particle that is located at
position $x$ at time $t$. From Eq.(\ref{psinu}) they are related by
\beq
p_x(v) = t \, p_x(q) \hspace{0.8cm} \mbox{and} \hspace{0.8cm} q = x - v t ,
\label{pvpq}
\eeq
since $q(x,t)$ is well defined for any $x$ except
over a set of zero measure in Eulerian space associated with shocks
\cite{She1992}.
As in \cite{Valageas2008}, we first consider the cumulative probability,
$p_x(0\leq q'\leq q)$, that the Lagrangian coordinate $q'(x,t)$
is within the range $0\leq q'\leq q$. From the geometrical construction 
(\ref{paraboladef}), this is the integral over the parabola height $c$
of the bivariate probability distribution, 
$p_x(0\leq q'\leq q,c)\dd c$, that the first-contact point of the potential 
$\psi_0$ with the family of downward parabolas $\cP_{x,c}$, with $c$ increasing 
from $-\infty$, occurs at an abscissa $q'$ in the range $0\leq q'\leq q$, with a 
parabola of height between $c$ and $c+\dd c$. 
In terms of the propagator $K_{x,c}$ introduced in Eq.(\ref{KG1}) this probability
density reads as
\beq
p_x(0\leq q'\leq q,c)\dd c = \lim_{q_+\rightarrow+\infty} 
\int \dd\psi\dd v \dd\psi_+\dd v_+
\, [ K_{x,c}(0,0,0;q,\psi,v) - K_{x,c+\dd c}(0,0,0;q,\psi,v) ] 
\, K_{x,c}(q,\psi,v;q_+,\psi_+,v_+) .
\label{pxcqF1}
\eeq
Here we used the Markovian character of the process $q\mapsto\{\psi,v\}$,
which allows us to factorize the probability $p_x(0\leq q'\leq q,c)\dd c$ into 
two terms, which correspond to the probabilities that
i) $\psi_0$ stays above $\cP_{x,c}$, but does not everywhere remain above 
$\cP_{x,c+\dd c}$, over the range $0\leq q'\leq q$, while reaching an arbitrary 
value $\{\psi,v\}$ at $q$, over which we will integrate, and ii) $\psi_0$ 
stays above $\cP_{x,c}$ for $q' > q$. 
We show in Fig.~\ref{figP1} the geometrical 
interpretation of Eq.(\ref{pxcqF1}) for a case with $x>0$ (we did not try to draw 
on the right side an actual Brownian curve $\psi_0(q)$ which has no finite 
second derivative).

The constraint associated with the left part of the potential $\psi_0$ at $q<0$
merely translates into an upper bound for the parabola height $c$.
Thus, if $x>0$, we must integrate over $c$ up to the value $c_*=x^2/(2t)$ where
the parabola $\cP_{x,c_*}$ runs through the origin $\{0,\psi_0(0)=0\}$. Indeed, it is
clear that all points on the negative axis $\{q<0,\psi_0(q)=0\}$ are still located
above $\cP_{x,c_*}$ hence they cannot be the minimum associated with (\ref{psinu}).
In fact, we can note that for any $x>0$ first contact always 
occurs before reaching $c_*$ because the initial potential has a zero derivative
at $q=0$ ($v_0(0)=0$). This also means that no rarefaction interval opens at $x=0^+$
(nor at any other location, see \cite{Bertoin1998}).  
On the other hand, for $x<0$ we must clearly integrate up to $c=0$.
Moreover, if the first contact is only reached at $q=x$ for $c=0$, it means that no 
particles from the right part have reached the position $x$ yet.
(For the case of two-sided Brownian initial velocity one would need to
add a third factor of the form $K_{x,c}(0,0,0;q_-,\psi_-,v_-)$ in Eq.(\ref{pxcqF1})
to take into account the left part of $\psi_0$, see \cite{Valageas2008}.) 

Using the expressions given in Appendix A, as well as the results of appendices A 
and B of \cite{Valageas2008}, we obtain from Eq.(\ref{pxcqF1}) for $x \geq 0$, 
after integration over $c$, the probability density
\beq
X \geq 0 : \;\; P_X(Q) = \inta \frac{\dd s}{2\pi\ii} \, e^{(s-1)Q} \, s^{-1/4} \,
e^{-(\sqrt{s}-1)2X} \;\;\; \mbox{over} \;\;\; Q\geq 0 ,
\label{PXpF} 
\eeq
which we expressed in terms of the dimensionless variables (\ref{QXdef}). Of course,
the distribution vanishes over $Q<0$, since particles from the left side cannot
travel to the right side $X\geq 0$.

In particular, at the origin $X=0$ this yields
\beq
Q \geq 0 : \;\;P_0(Q) = \frac{1}{\Gamma[1/4]} \, Q^{-3/4} \, e^{-Q} , \;\;
\mbox{whence} \;\;  P_0(V) = \frac{1}{\Gamma[1/4]} \, (-V)^{-3/4} \, e^{V}  
\;\; \mbox{with} \;\;  V\leq 0 ,
\label{PX0F} 
\eeq
where we used $V=-Q$ for $X=0$. The probability vanishes for $Q<0$ and $V>0$, as 
particles cannot come from the left side. Thus we recover the results of 
\citet{Bertoin1998}, who obtained Eqs.(\ref{PXpF})-(\ref{PX0F}) from 
probabilistic tools. The distribution of the time increments of $q(0,t)$,
i.e. of $q(0,t_2)-q(0,t_1)$, was obtained in \cite{Bertoin2001}.
We can note that Eq.(\ref{PX0F}) is identical to the 
large-velocity tail of the distribution obtained at $x=0$ for the case of 
two-sided Brownian initial conditions \cite{Valageas2008}. Hence, for rare events 
the tail of the distribution does not strongly depend on the initial conditions on 
the opposite side of the origin.

On the other hand, at large $X$ we obtain for fixed velocity, $V=X-Q$,
\beq
X \rightarrow +\infty : P_X(V) \sim \frac{e^{-V^2/X}}{\sqrt{\pi X}} .
\label{PXinfVF}
\eeq
Here we used the relationship $P_X(V)=P_X(Q=X-V)$ between the probability 
distributions of the velocity $V$ and of the Lagrangian coordinate $Q$, and the
explicit expression (\ref{PXpF}).
Thus, as for the two-sided case, we recover at leading order the initial Gaussian
distribution on large scales, here at $x\rightarrow +\infty$.
This is related to the ``principle of permanence of large eddies'' encountered in
the hydrodynamical context \cite{Gurbatov1997}, that holds for more general energy 
spectra, $E_0(k) \propto k^n$, with $n<1$. This states that regions of size
$x \gg L(t)$, where $L(t)$ is the integral scale of turbulence (here 
$L(t)\propto \gam^2=2Dt^2$, see the scalings (\ref{QXdef})), have not been strongly
distorted by smaller scale motions yet (since the relative distance between particles
has changed by an amount of order $L(t)$ at time $t$).
Thus, as checked in numerical simulations
\cite{Aurell1993,Gurbatov1999}, the stability of large-scale structures is not
only a statistical property but actually holds on an individual basis, that is 
for each random realization of the velocity field.
The properties of the velocity and Lagrangian increments on the right part of the 
system, $x>0$, were already obtained in \cite{Bertoin1998} and are also identical
to those obtained far from the origin in \cite{Valageas2008} for two-sided
initial conditions. 
In particular, it can be seen that the $n$-point distributions 
$p_{x_1,..,x_n}(q_1,..,q_n)$ factorize as 
$p_{x_1}(q_1) \bp_{x_{2,1}}(q_{2,1}) .. \bp_{x_{n,n-1}}(q_{n,n-1})$, where we note
$x_{i,i-1}=x_i-x_{i-1}$ and $q_{i,i-1}=q_i-q_{i-1}$ the relative distances, for
$x_1<..<x_n$ and $q_1<..<q_n$. 
Thus, the increments of the inverse Lagrangian map, $x\mapsto q$, are independent 
and have a simple distribution, which is given by the expression (\ref{PXpF}) 
without the factor $s^{-1/4}$. Then, over $x\geq 0$ the properties of the 
density field $\rho(x,t)$ are identical to those described in detail in 
\cite{Valageas2008} far from the origin, for the case of two-sided Brownian 
initial conditions.

On the left side, $x<0$, we must integrate Eq.(\ref{pxcqF1}) over $c$ up to $c=0$, 
as explained above. This yields
\beq
X \leq 0 : \;\; P_X(Q) = e^{2X} \inta \frac{\dd s}{2\pi\ii} \, e^{(s-1)Q}
\int_0^{\infty} \frac{\dd\nu}{\sqrt{\pi}} 3 \nu^{-3/2} 
e^{-\frac{2}{3}s^{3/2}\nu^{-3}-\nu^3X^2} \Ai\left[-\nu 2X +\frac{s}{\nu^2}\right] 
\;\;\; \mbox{over} \;\;\; Q \geq 0 .
\label{PXmF} 
\eeq
We can note that from Eqs.(\ref{PXpF}), (\ref{PXmF}), the tails at large $Q$ and 
$V$ read as
\beq
Q \rightarrow +\infty : \;\;  P_X(Q) \sim \frac{e^{2X}}{\Gamma[1/4]} \, 
Q^{-3/4} \, e^{-Q} , \;\; \mbox{and} , \;\; V\rightarrow-\infty : \;\;
P_X(V) \sim \frac{e^X}{\Gamma[1/4]} \, (-V)^{-3/4} \, e^V ,
\label{PXQVinfF}
\eeq
which hold for both $X\geq 0$ and $X\leq 0$. Thus, at any finite $X$ the tail of 
the velocity distribution simply follows the exponential decay obtained at $X=0$ 
in (\ref{PX0F}), multiplied by a prefactor $e^X$.

\begin{figure}
\begin{center}
\epsfxsize=8 cm \epsfysize=6.2 cm {\epsfbox{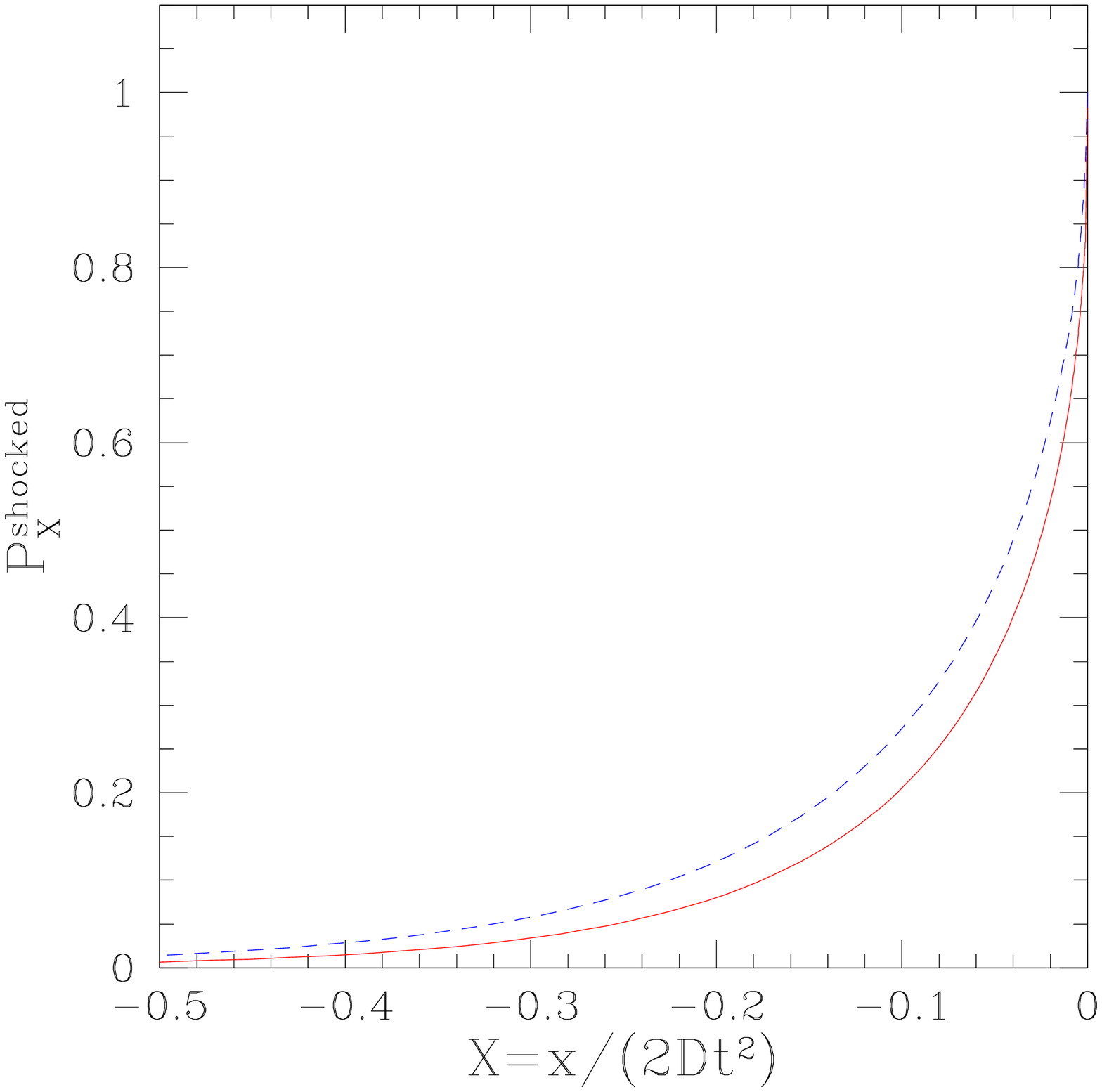}}
\epsfxsize=8 cm \epsfysize=6.2 cm {\epsfbox{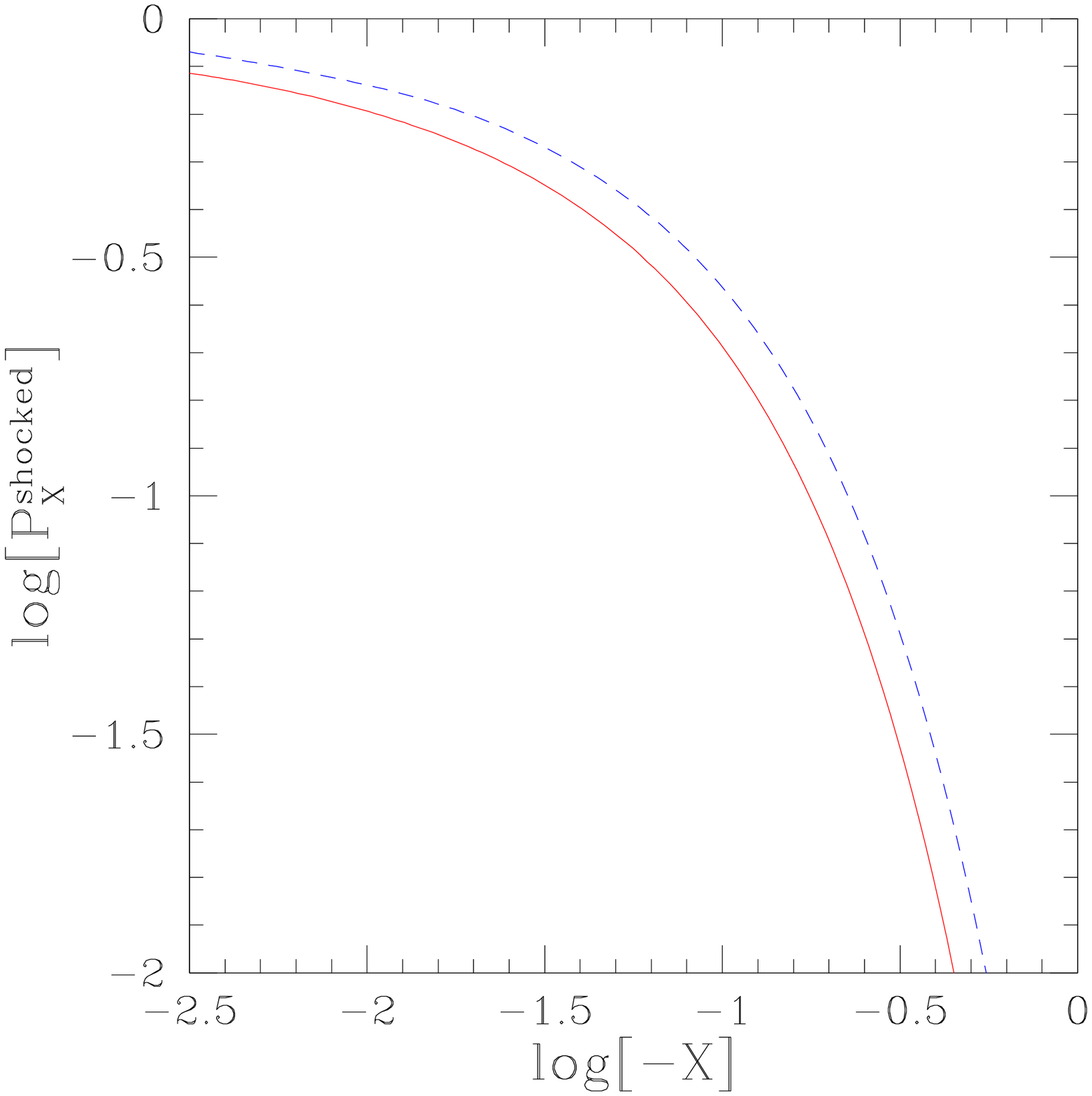}}
\end{center}
\caption{{\it Left panel:} The probability $P_X^{\rm shocked}$ that particles 
from the right side have already reached the position $X$ on the left side.
We show our results for the filled case ``$F$'' (solid line, Eq.(\ref{PXshockedF}))
and the empty case ``$E$'' (dashed line, Eq.(\ref{PXshockedE})).
{\it Right panel:} Same as left panel but on a logarithmic scale.}
\label{figPshock}
\end{figure}

The distribution (\ref{PXmF}) corresponds to realizations where some particles 
coming from the right side have already passed by position $x$. Thus, integrating 
Eq.(\ref{PXmF}) over $Q\geq 0$ gives the probability, $P_X^{\rm shocked}$, that 
matter coming from the right side has already passed by the position $X<0$. This 
yields
\beq
X \leq 0 : \;\; P_X^{\rm shocked} = e^{2X} \int_0^{\infty} 
\frac{\dd\nu}{\sqrt{\pi}} 3 \nu^{-3/2} e^{-\frac{2}{3}\nu^{-3}-\nu^3X^2} 
\Ai\left[-\nu 2X +\frac{1}{\nu^2}\right] .
\label{PXshockedF} 
\eeq
This could also be obtained directly by computing as in (\ref{pxcqF1}) the 
probability that the curve $\psi_0$ goes below the parabola $\cP_{x,0}$ at some 
point $q\geq 0$. Then, to the contribution (\ref{PXmF}) we must add the 
contribution $P_X^{\rm not-shocked} \delta(Q-X)$, with 
$P_X^{\rm not-shocked}=1-P_X^{\rm shocked}$, that corresponds to realizations where 
no particles from the right side have already passed by position $x$, so that the 
medium has remained at rest at $x$ until time $t$.
Finally, Eq.(\ref{PXshockedF}) gives the asymptotic behaviors
\beq
X \rightarrow 0^- : \;\; P_X^{\rm shocked} \sim 1 - 
\frac{2^{4/3} \, 3^{1/3}}{\Gamma[1/3]} \, (-X)^{1/3} , \;\;\;\;
X \rightarrow -\infty : \;\; P_X^{\rm shocked} \sim \frac{3^{4/3}}{16} 
\, \sqrt{\frac{5}{-2\pi X}} \,\, e^{64X/9} .
\label{PXshockedasympF}
\eeq
We can check that $P_0^{\rm shocked} =1$ at $X=0$, while it shows an exponential 
tail at large negative $X$. Since all quantities can be expressed in terms of the 
scaled variables (\ref{QXdef}), as we noticed from (\ref{scalings}) and can be 
checked in the results above, this exponential decay can be obtained from simple 
scaling arguments. Thus, for particle $q>0$ to reach the Eulerian position $x<0$ 
at time $t$, we can expect its initial velocity to be of order $v_0 \sim (x-q)/t$. 
Since the initial velocity is Gaussian, of variance given by Eq.(\ref{Ddef}), this 
corresponds to a probability of order $e^{-v_0^2/(2Dq)}=e^{-(x-q)^2/(2Dt^2q)}$. 
This is maximum at $q=-x$, which gives a weight $e^{4x/(2Dt^2)}=e^{4X}$. Hence we 
recover an exponential over $X$. Of course, such a reasoning cannot give the 
numerical factor within the exponential. 
We show in Fig.~\ref{figPshock} our results for the probability $P_X^{\rm shocked}$,
given by Eq.(\ref{PXshockedF}), for the present case ``$F$'' where the left side 
is initially filled by particles at rest (solid line) (we also display for 
comparison the result associated with the empty case ``$E$'', that we shall
obtain below in Eq.(\ref{PXshockedE}) (dashed line)). We clearly see the fast
decline with larger $|X|$ obtained in Eq.(\ref{PXshockedasympF}).

\subsection{Mean density profile and mean current}
\label{subsec:density-profile}

From Eq.(\ref{PXpF}) we also obtain the mean Lagrangian coordinate $\lag Q(X)\rag$,
and velocity $\lag V(X)\rag$, at Eulerian location $X\geq 0$ on the right side, as
\beq
X\geq 0 : \;\; \lag Q(X)\rag= X+\frac{1}{4} , \;\; \lag V(X) \rag= -\frac{1}{4} ,
\;\;\; \mbox{whence} \;\;\;  \lag q(x)\rag= x+\frac{Dt^2}{2} , \;\; 
\lag v(x) \rag= -\frac{Dt}{2} .
\label{meanQVpF}
\eeq
As expected, since particles gradually leak into the left side the mean velocity is 
negative, and particles that occupy the Eulerian position $x$ come from increasingly 
far regions on the right as time increases. We also obtain the mass, $m(<x)$, of 
excited particles (i.e. with initial Brownian velocity and which were initially 
located at $q\geq 0$) that are located to the left of the Eulerian position $x$ by 
noting that it is given by $m(<x)=\rho_0 q(x,t)$ since particles do not cross
each other. This yields from Eq.(\ref{meanQVpF})
\beq
x\geq 0 : \;\; \lag m(<x)\rag= \rho_0 (x+Dt^2/2) , \;\;\; \mbox{whence} \;\;
\lag \rho(x)\rag = \rho_0 \;\; \mbox{and} \;\; \lag j(x) \rag = - \rho_0 Dt ,
\;\;\; \mbox{with} \;\; \rho=\frac{\pl m}{\pl x} \;\; \mbox{and} \;\;
j= - \frac{\pl m}{\pl t} ,
\label{meanMrhojpF}
\eeq
where we introduced the density $\rho(x)$ and the current $j(x)$ at position $x$ of 
excited particles. Therefore, we obtain a uniform mean flow from the right,
with the mean current $\lag j(x)\rag = -\rho_0Dt$, while the mean density remains 
equal to $\rho_0$. Note that $\lag j(x)\rag = 2 \lag \rho(x)\rag  \lag v(x) \rag$, 
which implies that the fluctuations of the density and velocity are correlated.
In fact, the velocity field $v(x,t)$ is associated with Eulerian regular points,
since shocks have a zero measure, but the flow of matter is associated with shocks 
since all the mass of excited particles is contained within shocks 
\cite{She1992,Sinai1992,Bertoin2001,Valageas2008}.
Therefore, it is not surprising to find that $\lag j(x)\rag \neq \lag \rho(x)\rag 
\lag v(x) \rag$, since these quantities probe different aspects of the dynamics.

Thus, even though particles keep escaping into the left part $x<0$, particles coming 
from the right semi-infinite axis keep replenishing the system and manage to 
maintain a constant mean density $\rho_0$ over $x>0$, through the mean uniform 
current $-\rho_0  Dt$ that grows linearly with time. The linear growth with time
of the mean velocity and current is due to the fact that at later times particles 
coming from more distant regions have been able to reach the boundary $x=0$. 
Again, this exponent can be 
obtained from simple scaling arguments. Thus, at time $t$ we can expect to see at 
the boundary, $x=0$, particles coming from a distance $q$ with an initial velocity 
of order $v_0 \sim -q/t$. Since the initial velocity scales as $q^{1/2}$, see 
(\ref{Ddef}), this gives $q \sim t^2$ and $v_0 \sim -t$, whence $v \sim -t$, 
assuming that matter flows through the boundary in a well-ordered fashion, so that 
the velocity of these particles has not been significantly damped by nearer 
lower-velocity particles, as the latter have already escaped into the left side.

\begin{figure}
\begin{center}
\epsfxsize=8 cm \epsfysize=6.2 cm {\epsfbox{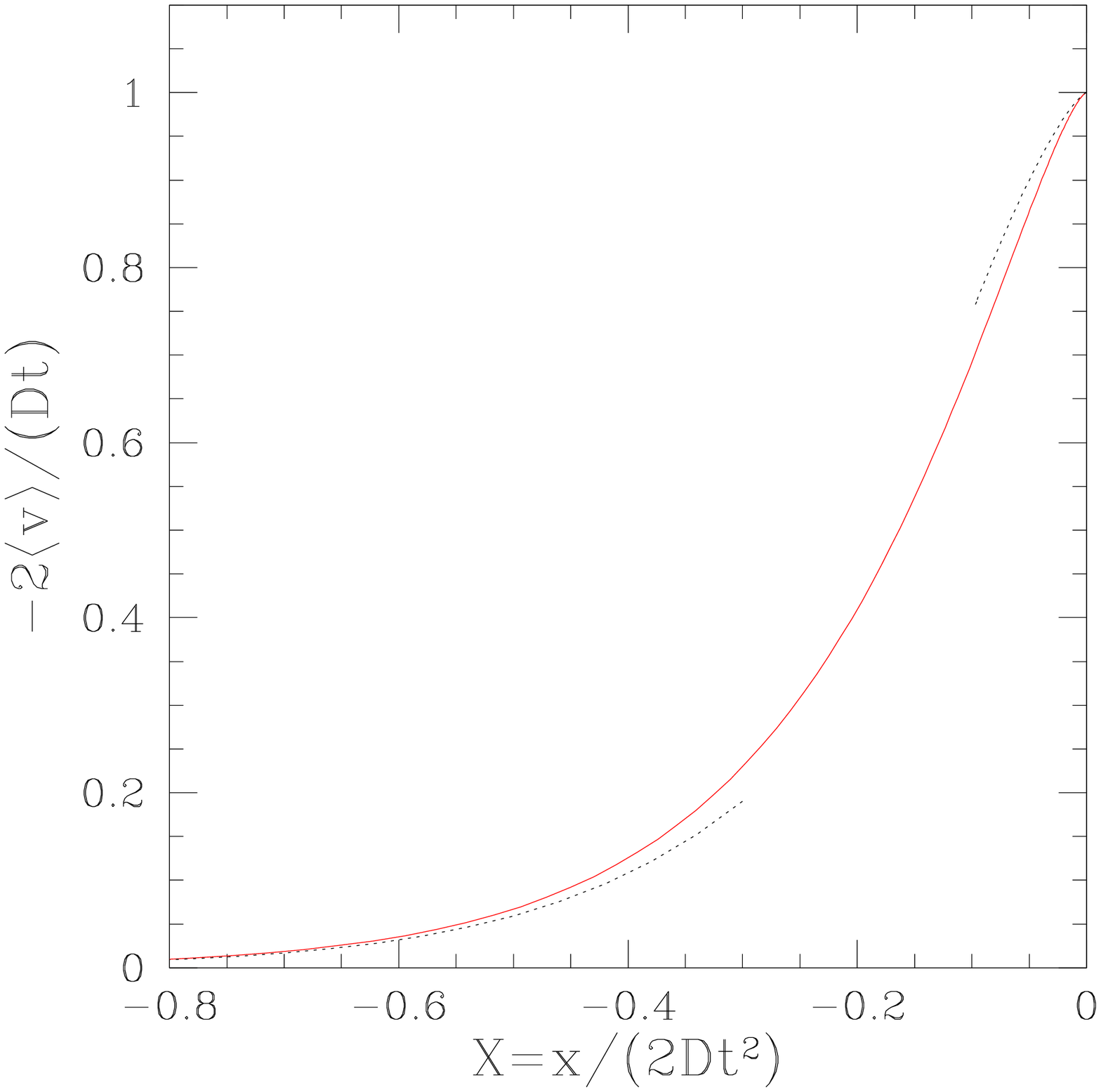}}
\epsfxsize=8 cm \epsfysize=6.2 cm {\epsfbox{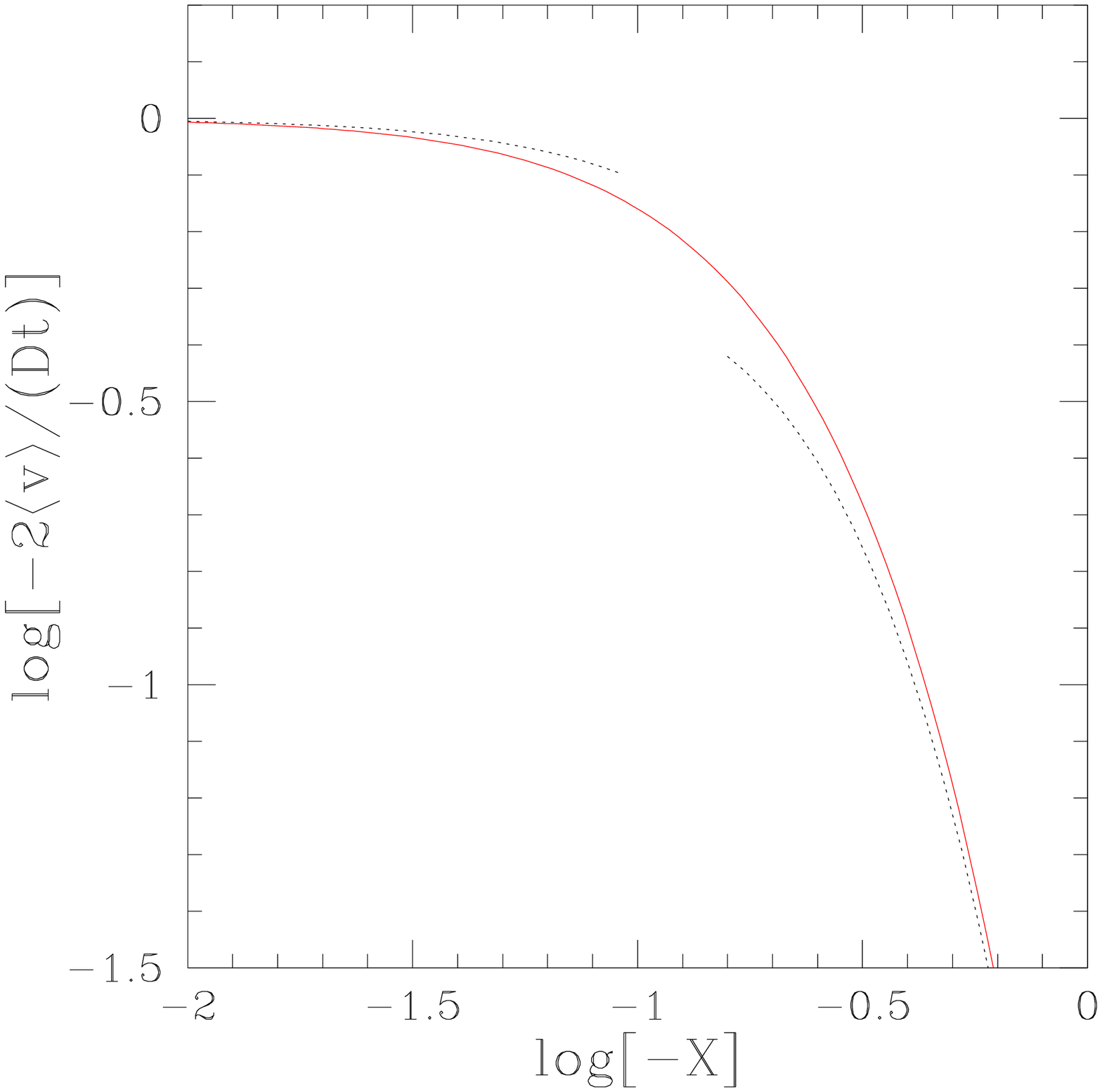}}
\end{center}
\caption{{\it Left panel:} The mean velocity profile $-2\lag v\rag/(Dt)$ on
the left side, $x\leq 0$ (on the right side the mean velocity is constant and
equal to $-Dt/2$). 
We show our results for the filled case ``$F$'' (solid line, Eq.(\ref{meanQmF}))
where the mean velocity is always meaningful. The dotted lines are the asymptotic 
behaviors (\ref{VasympF}).
{\it Right panel:} Same as left panel but on a logarithmic scale.}
\label{figvmean}
\end{figure}

On the left side, $X\leq 0$, we obtain from Eqs.(\ref{PXmF})-(\ref{PXshockedF})
the mean Lagrangian coordinate $\lag Q(X)\rag$, and velocity $\lag V(X)\rag$, as
\beq
X\leq 0 : \;\; \lag V(X)\rag= - e^{2X} \int_0^{\infty} \frac{\dd\nu}{\sqrt{\pi}} 3 
\nu^{-3/2} e^{-\frac{2}{3}\nu^{-3}-\nu^3X^2} \left[ (-X+\nu^{-3}) \, 
\Ai\left(-\nu 2X +\nu^{-2}\right) - \nu^{-2}\, \Ai\,'\left(-\nu 2X +\nu^{-2}\right) 
\right] ,
\label{meanQmF}
\eeq
and $\lag Q(X)\rag=X-\lag V(X)\rag$. This gives for the mean velocity the asymptotic 
behaviors
\beq
X \rightarrow 0^- : \;\;  \lag V(X)\rag \sim -\frac{1}{4}
+\frac{2^{4/3}3^{1/3}}{\Gamma[1/3]} (-X)^{4/3} , \;\;\; X \rightarrow -\infty : 
\;\; \lag V(X)\rag \sim -\frac{3}{4} \sqrt{\frac{-3X}{\pi}} \, e^{64X/9} .
\label{VasympF}
\eeq
We show our results for the mean velocity profile in Fig.~\ref{figvmean}.
We can check that $|\lag V(X)\rag|$ shows a monotonic
decrease for larger $|X|$ on the left side. As expected, we recover the same 
asymptotic exponential decay as for the probability $P_X^{\rm shocked}$ obtained 
in Eq.(\ref{PXshockedasympF}).

\begin{figure}
\begin{center}
\epsfxsize=8 cm \epsfysize=6.2 cm {\epsfbox{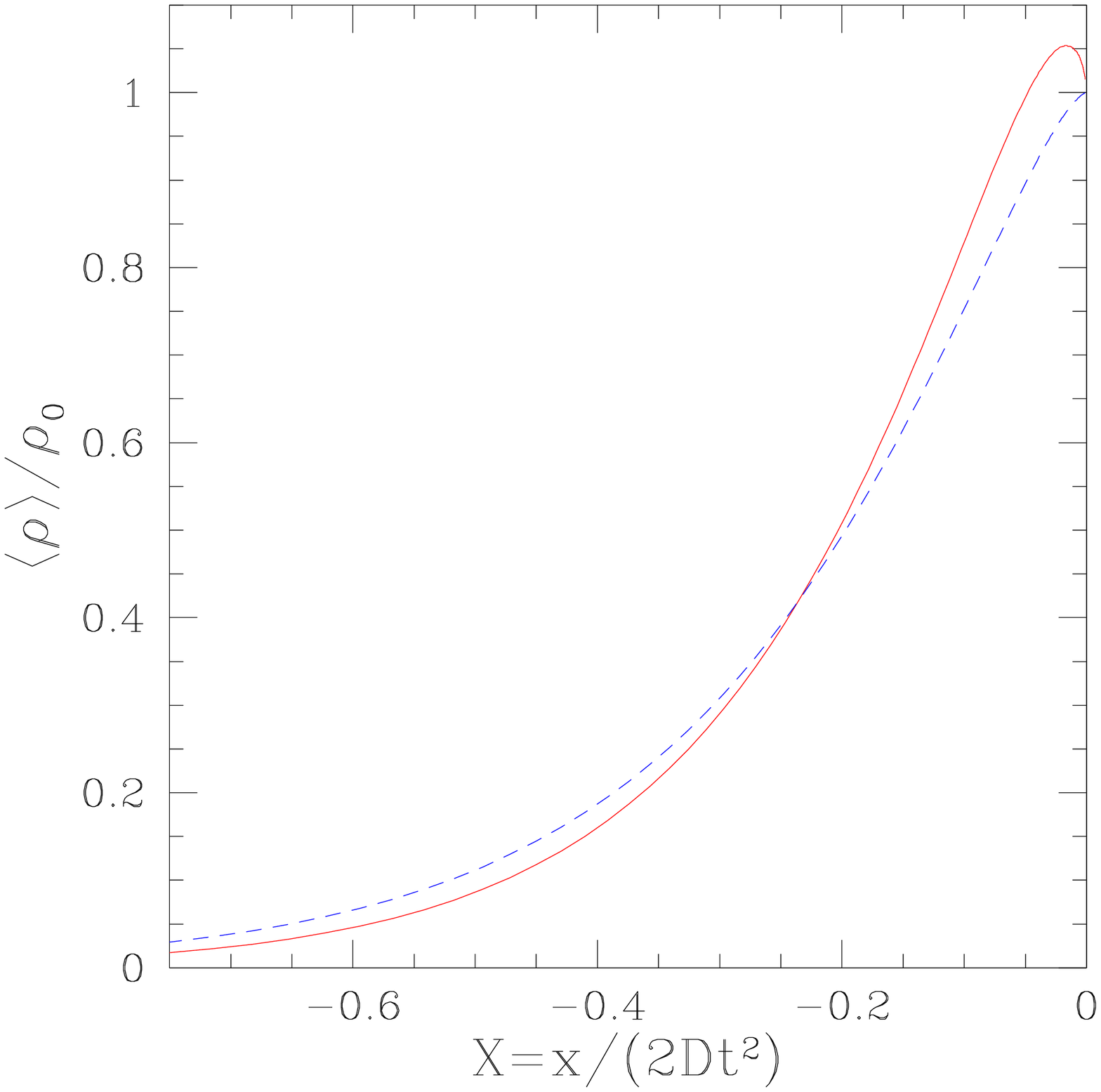}}
\epsfxsize=8 cm \epsfysize=6.2 cm {\epsfbox{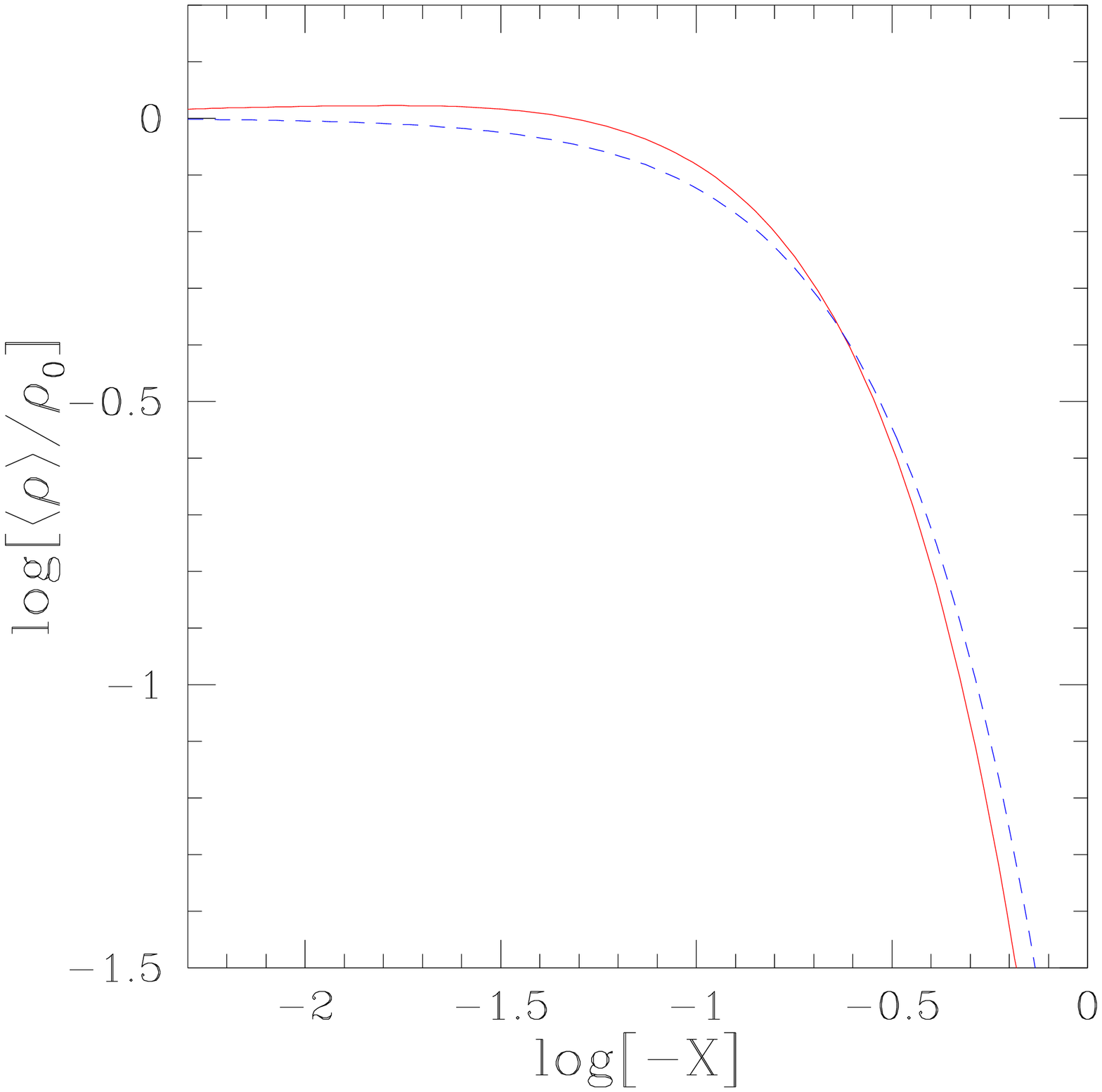}}
\end{center}
\caption{{\it Left panel:} The mean density profile $\lag\rho\rag/\rho_0$ on
the left side, $x\leq 0$ (on the right side the mean density is constant and
equal to $\rho_0$). 
We show our results for the filled case ``$F$'' (solid line, Eq.(\ref{meanrhomF}))
and the empty case ``$E$'' (dashed line, Eq.(\ref{meanrhomE})).
{\it Right panel:} Same as left panel but on a logarithmic scale.}
\label{figrho}
\end{figure}

\begin{figure}
\begin{center}
\epsfxsize=8 cm \epsfysize=6.2 cm {\epsfbox{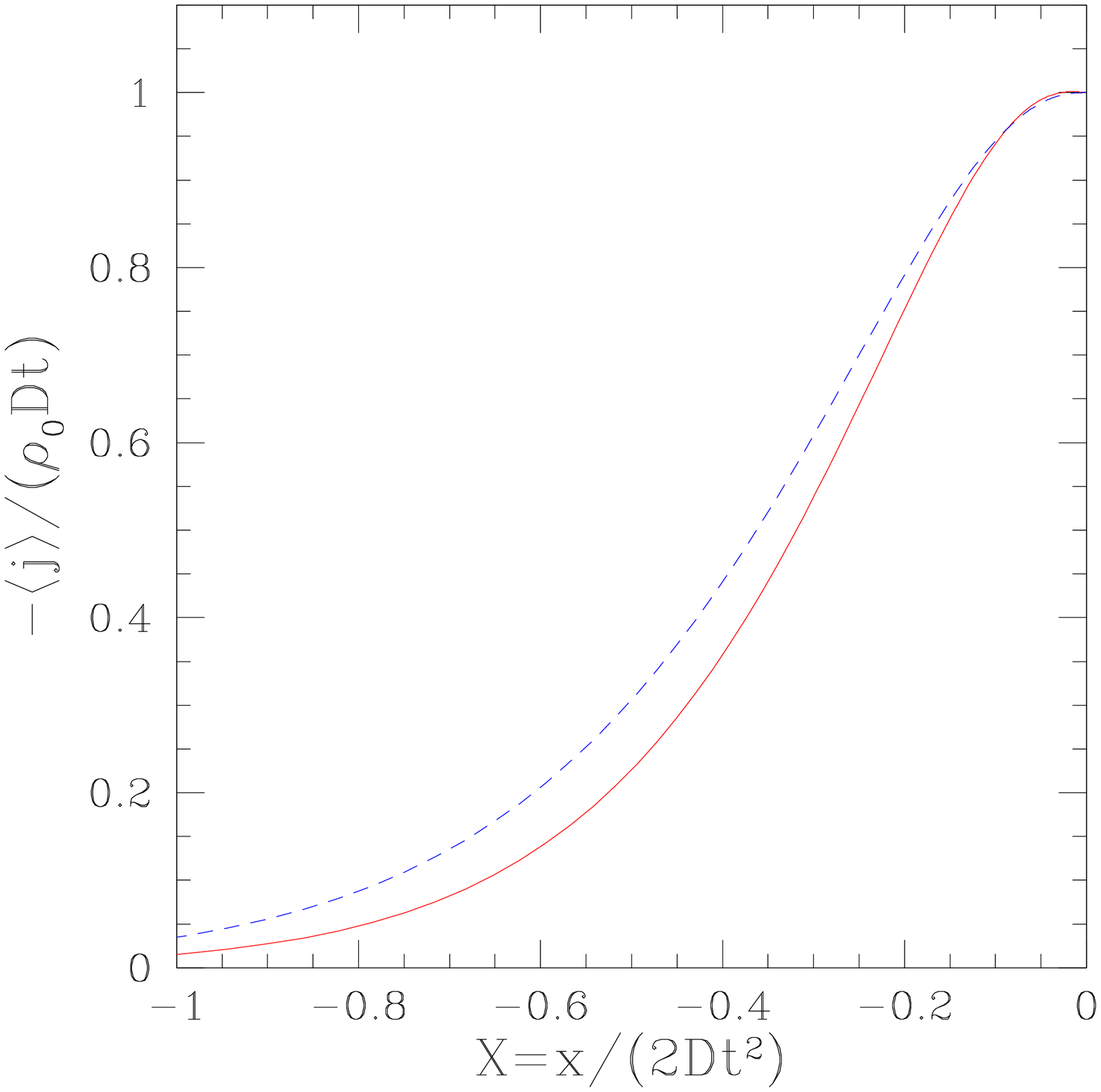}}
\epsfxsize=8 cm \epsfysize=6.2 cm {\epsfbox{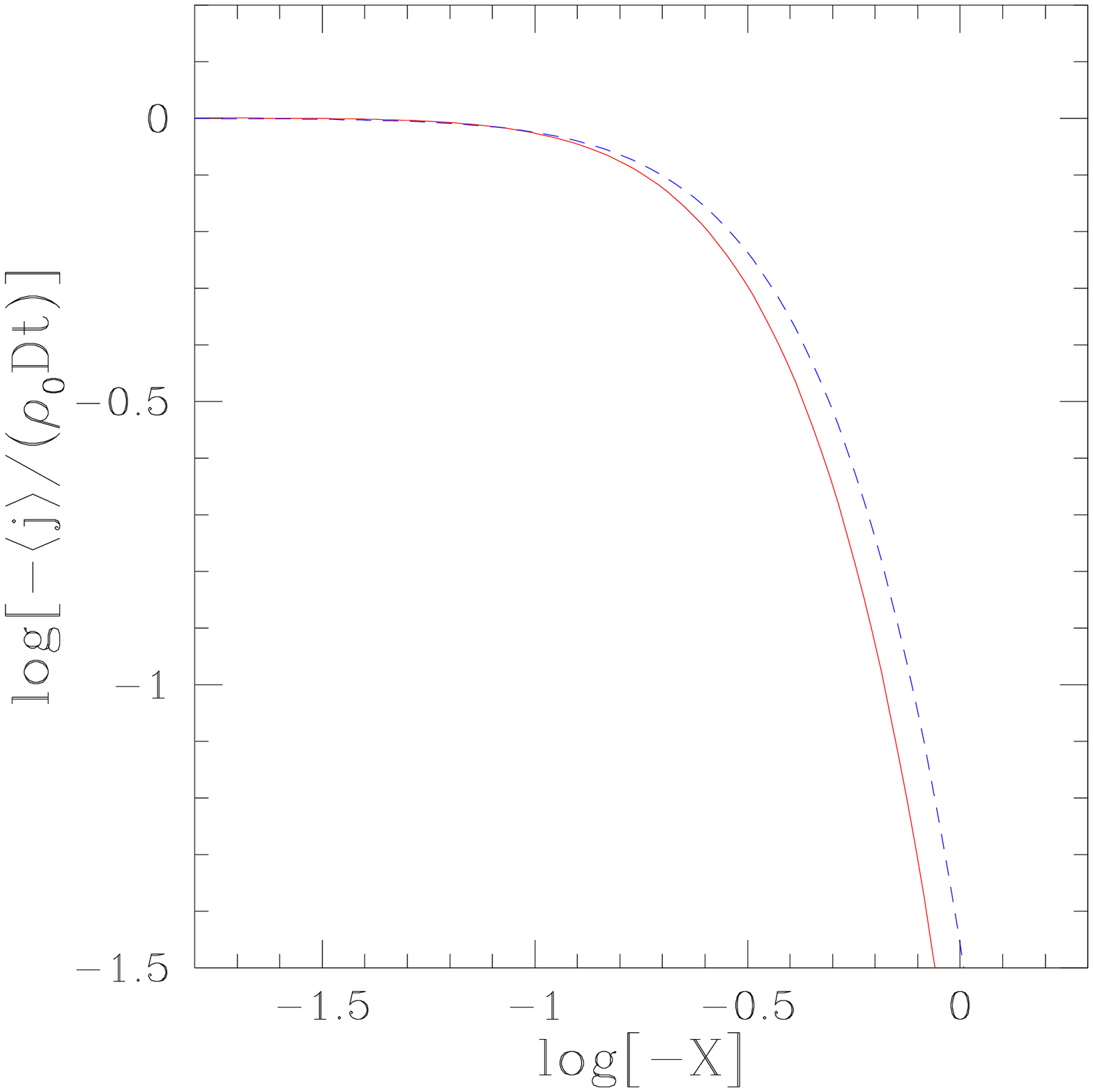}}
\end{center}
\caption{{\it Left panel:} The mean current profile $-\lag j\rag/(\rho_0 Dt)$ on
the left side, $x\leq 0$ (on the right side the mean current is constant and
equal to $-\rho_0 Dt$). 
We show our results for the filled case ``$F$'' (solid line, Eq.(\ref{meanjmF}))
and the empty case ``$E$'' (dashed line, Eq.(\ref{meanjmE})).
{\it Right panel:} Same as left panel but on a logarithmic scale.}
\label{figjmean}
\end{figure}

The average (\ref{meanQmF}) takes into account both the contributions 
(\ref{PXmF}) and $P_X^{\rm not-shocked} \delta(Q-X)$. However, since only the 
part (\ref{PXmF}) contributes to the mass of excited particles located to 
the left of $X$, we no longer have $\lag M(<X) \rag=\lag Q(X)\rag$, as was the case 
for $X\geq 0$ in (\ref{meanMrhojpF}), where we introduced the dimensionless mass 
$M$ defined by
\beq
M= \frac{m}{\rho_0 \gam^2} , \;\; \mbox{whence} \;\; 
\lag\rho\rag= \frac{\pl \lag m\rag}{\pl x} = \rho_0 \frac{\pl \lag M\rag}{\pl X} \;\; 
\mbox{and} \;\; \lag j\rag = - \frac{\pl \lag m\rag}{\pl t} = \frac{2}{t} \rho_0 
\gam^2 \left(X \frac{\pl \lag M\rag}{\pl X}-\lag M\rag\right) .
\label{Mdef}
\eeq
In the last two relations in (\ref{Mdef}) we used the property that $\lag M(<X)\rag$ 
only depends on the reduced variable $X$. Integrating over the contribution 
(\ref{PXmF}) gives
\beq
X\leq 0 : \;\; \lag M(<X)\rag= e^{2X} \int_0^{\infty} \frac{\dd\nu}{\sqrt{\pi}} 
3 \nu^{-3/2} e^{-\frac{2}{3}\nu^{-3}-\nu^3X^2} \left[ \nu^{-3} \, 
\Ai\left(-\nu 2X +\nu^{-2}\right) - \nu^{-2}\, \Ai\,'\left(-\nu 2X +\nu^{-2}\right) 
\right]  ,
\label{meanMmF}
\eeq
as well as
\beq
x\leq 0 : \;\; \lag \rho(x)\rag= \rho_0 \, e^{2X} \int_0^{\infty} 
\frac{\dd\nu}{\sqrt{\pi}} 3 \nu^{-3/2} e^{-\frac{2}{3}\nu^{-3}-\nu^3X^2} 
\left[ (4\nu^{-3}-6X) \, \Ai\left(-\nu 2X +\nu^{-2}\right) - (4\nu^{-2}-2\nu X) \, 
\Ai\,'\left(-\nu 2X +\nu^{-2}\right) \right] ,
\label{meanrhomF}
\eeq
and
\beqa
x\leq 0 & : & \lag j(x)\rag= -4\rho_0 D t \, e^{2X} \int_0^{\infty} 
\frac{\dd\nu}{\sqrt{\pi}} 3 \nu^{-3/2} e^{-\frac{2}{3}\nu^{-3}-\nu^3X^2} 
\left[ (\nu^{-3}-4\nu^{-3}X+6X^2) \, \Ai\left(-\nu 2X +\nu^{-2}\right) \right.
\nonumber \\
&& \hspace{4cm} \left. - (\nu^{-2}-4\nu^{-2}X+2\nu X^2) \, 
\Ai\,'\left(-\nu 2X +\nu^{-2}\right) \right] .
\label{meanjmF}
\eeqa
We show in Figs.~\ref{figrho}, \ref{figjmean}, our results for the mean density and 
current.
Then, we can check that the mean density and current are continuous at the
boundary $x=0$, and we obtain the asymptotic behaviors
\beqa
X \rightarrow 0^- & : & \lag M(<X)\rag \sim \frac{1}{4} + X -
 \frac{2^{5/3}3^{2/3}}{5\Gamma[2/3]} (-X)^{5/3} , \;\;\;\; \lag \rho(x)\rag \sim 
\rho_0 \left[ 1 + \frac{2^{5/3}3^{-1/3}}{\Gamma[2/3]} (-X)^{2/3} \right] , 
\nonumber \\ 
&& \lag j(x)\rag \sim - \rho_0 D t \left[ 1 + \frac{2^{14/3}3^{-1/3}}{5\Gamma[2/3]} 
(-X)^{5/3} \right] .
\label{meanMrhojX0asympmF}
\eeqa

From Eq.(\ref{meanMrhojX0asympmF}), we can note that the amplitude of the mean 
density and current first increases to the left of the boundary $x=0$ (contrary to 
the mean velocity which showed a monotonic decrease into the left side). 
This is due to the dragging effect of the 
matter that was initially at rest on the left side, which slows down the leftmost
cluster associated with particles coming from the right side. Then, this 
deterministic friction leads to a transitory growth for the mean density and current
to the left of the boundary $x=0$. Figures~\ref{figrho} and \ref{figjmean} show that
although this feature can be clearly seen for the mean density (although it remains
modest, of order $5\%$), it is almost invisible for the current, in agreement with 
the higher power $(-X)^{5/3}$ instead of $(-X)^{2/3}$ in 
Eq.(\ref{meanMrhojX0asympmF}), which leads to a suppression by a factor $|X|$ at
small $X$.
As we shall check in section~\ref{subsec/E:density-profile} below, this feature 
disappears when the left side is initially empty.
The same behavior was obtained in \cite{Frachebourg2001} for the case of white-noise
initial velocity.
At large distance from the boundary we obtain the exponential decays
\beq
X \rightarrow -\infty : \;\;  \lag M(<X)\rag \sim \frac{9}{16} 
\sqrt{\frac{-3X}{\pi}} \, e^{64X/9} , \;\;\; \lag \rho(x)\rag \sim \rho_0 \, 4 
\sqrt{\frac{-3X}{\pi}} \, e^{64X/9} , \;\;\; \lag j(x)\rag \sim - \rho_0 D t \, 16 
\sqrt{\frac{3}{\pi}} \, (-X)^{3/2} \, e^{64X/9} .
\label{meanVMrhojasympmF}
\eeq
Again, the characteristic length scale is the reduced variable $X$
of (\ref{QXdef}), and the exponential decay is the same as the one obtained
in Eq.(\ref{PXshockedasympF}) for the shock probability $P_X^{\rm shocked}$.

\subsection{Lagrangian displacement field}
\label{subsec:Lagrangian_displacement}

We now consider the dynamics from a Lagrangian point of view. Thus, labelling the
particles by their initial position $q$ at time $t=0$, we follow their trajectory
$x(q,t)$. Since particles do not cross each other, the probability, $p_q(x'\geq x)$,
for the particle $q$ to be located to the right of the Eulerian position $x$ at time
$t$, is equal to the probability, $p_x(q'\leq q)$, for the Eulerian location $x$
to be ``occupied'' by particles that were initially to the left of particle $q$.
In terms of dimensionless variables, this gives for right-side particles, $q\geq 0$,
\beq
Q \geq 0 : \;\; P_Q(X'\geq X) = P_X(0\leq Q'\leq Q) = 
\inta \frac{\dd s}{2\pi\ii} \, e^{(s-1)Q} \, \frac{s^{-1/4}}{s-1} \,
e^{-(\sqrt{s}-1)2X} \;\;\; \mbox{over} \;\;\; X\geq 0 ,
\label{PQXpF} 
\eeq
where the integration contour runs to the right of the pole, i.e. $\Re(s)>1$, and
\beq
Q \geq 0 : \;\; P_Q(X'\geq X) = 1 - e^{2X} \inta \frac{\dd s}{2\pi\ii} \, 
\frac{e^{(s-1)Q}}{1-s} \int_0^{\infty} \frac{\dd\nu}{\sqrt{\pi}} 3 \nu^{-3/2} 
e^{-\frac{2}{3}s^{3/2}\nu^{-3}-\nu^3X^2} \Ai\left[-\nu 2X +\frac{s}{\nu^2}\right] 
\;\;\; \mbox{over} \;\;\; X\leq 0 ,
\label{PQXmF} 
\eeq
where the integration contour obeys $0<\Re(s)<1$.
Note that for $X\leq 0$ we must take into account both the probability, 
$1-P_X^{\rm shocked}$, that no particles from the right side have reached $X$ yet,
and the probability, $P_X(0\leq Q'\leq Q)$, that particles $Q'$ from the right side,
with $0\leq Q'\leq Q$, have already passed by point $X$.
Then, probability densities are obtained from Eqs.(\ref{PQXpF})-(\ref{PQXmF}) by
differentiating with respect to $X$.

\subsection{Leftmost cluster}
\label{subsec:Leftmost_cluster}

\begin{figure}
\begin{center}
\epsfxsize=8 cm \epsfysize=6.2 cm {\epsfbox{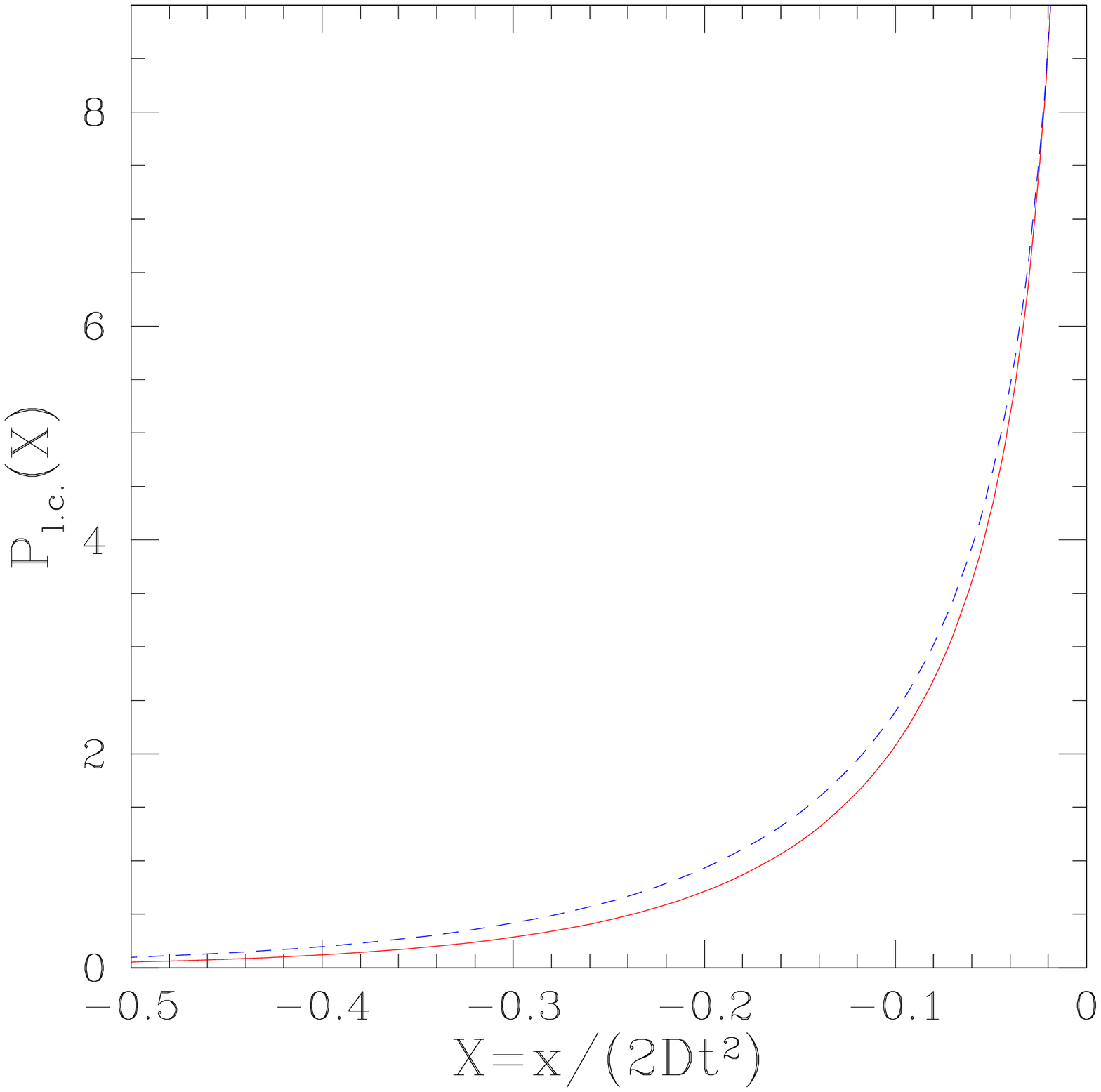}}
\epsfxsize=8 cm \epsfysize=6.2 cm {\epsfbox{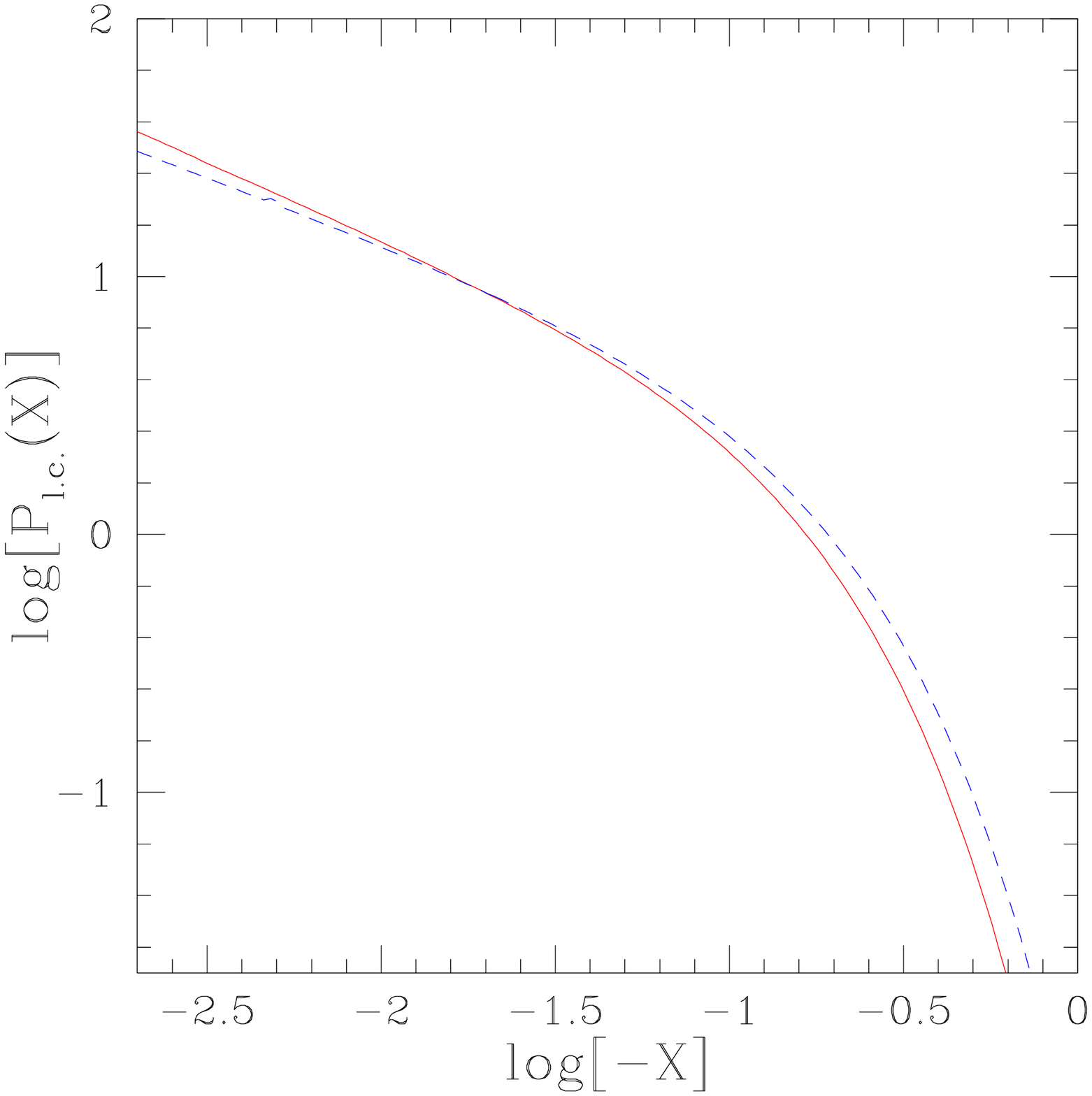}}
\end{center}
\caption{{\it Left panel:} The probability distribution $P_{\rm l.c.}(X)$ of the 
position $X$ of the leftmost cluster.
We show our results for the filled case ``$F$'' (solid line, Eq.(\ref{PlcXF}))
and the empty case ``$E$'' (dashed line, Eq.(\ref{PlcXE})).
{\it Right panel:} Same as left panel but on a logarithmic scale.}
\label{figPlcX}
\end{figure}

We can identify the Eulerian position, $x(0,t)$, of the particle that was initially 
located at the origin ($q=0$), as the position of the leftmost cluster (or ``leader'')
formed by excited particles that have escaped into the left side. 
In the present case, where 
these particles spread into a medium of uniform density that was initially at rest, 
this cluster also contains a mass $\rho_0 |x|$ of particles that were located 
in the interval $[x,0]$. It acts as a snow-plough while the conditional properties 
of the system to its right are no longer sensitive to the initial conditions on the 
left side.
From Eq.(\ref{PQXpF}) we can see that $P_0(X)$ vanishes for $X\geq 0$, which means 
that at any time $t>0$ the particle $q=0$ has almost surely already passed to the 
left side and the leftmost cluster has formed with a finite mass, in agreement with 
the results of \cite{Isozaki2006}. On the other hand, for $X< 0$ the probability 
density $P_{\rm l.c.}(X)$ of the leftmost cluster position reads as
\beq
X < 0 : \;\; P_{\rm l.c.}(X) = \frac{\dd}{\dd X} P_X^{\rm shocked} = 
\frac{\dd}{\dd X} \, e^{2X} \int_0^{\infty} \frac{\dd\nu}{\sqrt{\pi}} 3 \nu^{-3/2} 
e^{-\frac{2}{3}\nu^{-3}-\nu^3X^2} \Ai\left[-\nu 2X+\frac{1}{\nu^2}\right] .
\label{PlcXF}
\eeq
Here we used the obvious property 
$P_{\rm l.c.}(X'\geq X)=P_X^{\rm not-shocked}=1- P_X^{\rm shocked}$, which states that
the leftmost cluster is located to the right of point $X$ if, and only if, no
particles from the right side have reached this point yet. We can check that the 
result (\ref{PlcXF}) is identical to the one that would be obtained from 
Eq.(\ref{PQXmF}). Using Eq.(\ref{PXshockedasympF}) we obtain the asymptotic 
behaviors
\beq
X \rightarrow 0^- : \;\; P_{\rm l.c.}(X) \sim  \frac{2^{4/3} \, 3^{-2/3}}
{\Gamma[1/3]} (-X)^{-2/3} , \;\;\;\; X \rightarrow -\infty : \;\; P_{\rm l.c.}(X) 
\sim \frac{4}{3^{2/3}} \sqrt{\frac{5}{-2\pi X}} \,\, e^{64X/9} .
\label{PlcXasympF}
\eeq
We show in Fig.~\ref{figPlcX} our result (\ref{PlcXF}) for $P_{\rm l.c.}(X)$.
From Eq.(\ref{PlcXF}), we obtain after an integration by parts the mean position 
of the leftmost cluster as
\beq
\lag x_{\rm l.c.}(t)\rag = \lag X_{\rm l.c.}\rag 2D t^2 \;\;\;\; \mbox{with} 
\;\;\;\; \lag X_{\rm l.c.}\rag = - \int_{-\infty}^0 \dd X \, e^{2X} \int_0^{\infty}
\frac{\dd\nu}{\sqrt{\pi}} 3 \nu^{-3/2} e^{-\frac{2}{3}\nu^{-3}-\nu^3X^2} 
\Ai\left[-\nu 2X+\frac{1}{\nu^2}\right]  \simeq -0.06 .
\label{XlcF}
\eeq
Thus, the distance $|x_{\rm l.c.}(t)|$ from the origin scales with time as $t^2$. 
The leftmost cluster has a super-ballistic motion because it is constantly overtaken
by higher-velocity particles coming farther away from the right side which increase
its momentum. Since the growth of the mean $\lag x_{\rm l.c.}(t)\rag$ is set by the 
scaling variables (\ref{QXdef}) it can be understood from simple arguments,
as for the mean current obtained in (\ref{meanMrhojpF}). There, we have seen that
the scaling $v(0,t) \sim -t$ could be explained by the time needed for particles
at distance $q\sim t^2$ to reach the origin. Then, if the position of the leftmost
cluster is set by the latest particles that escaped into the left side we
expect $x_{\rm l.c.} \sim v t \sim -t^2$, which agrees with (\ref{XlcF}).

\begin{figure}
\begin{center}
\epsfxsize=8 cm \epsfysize=6.2 cm {\epsfbox{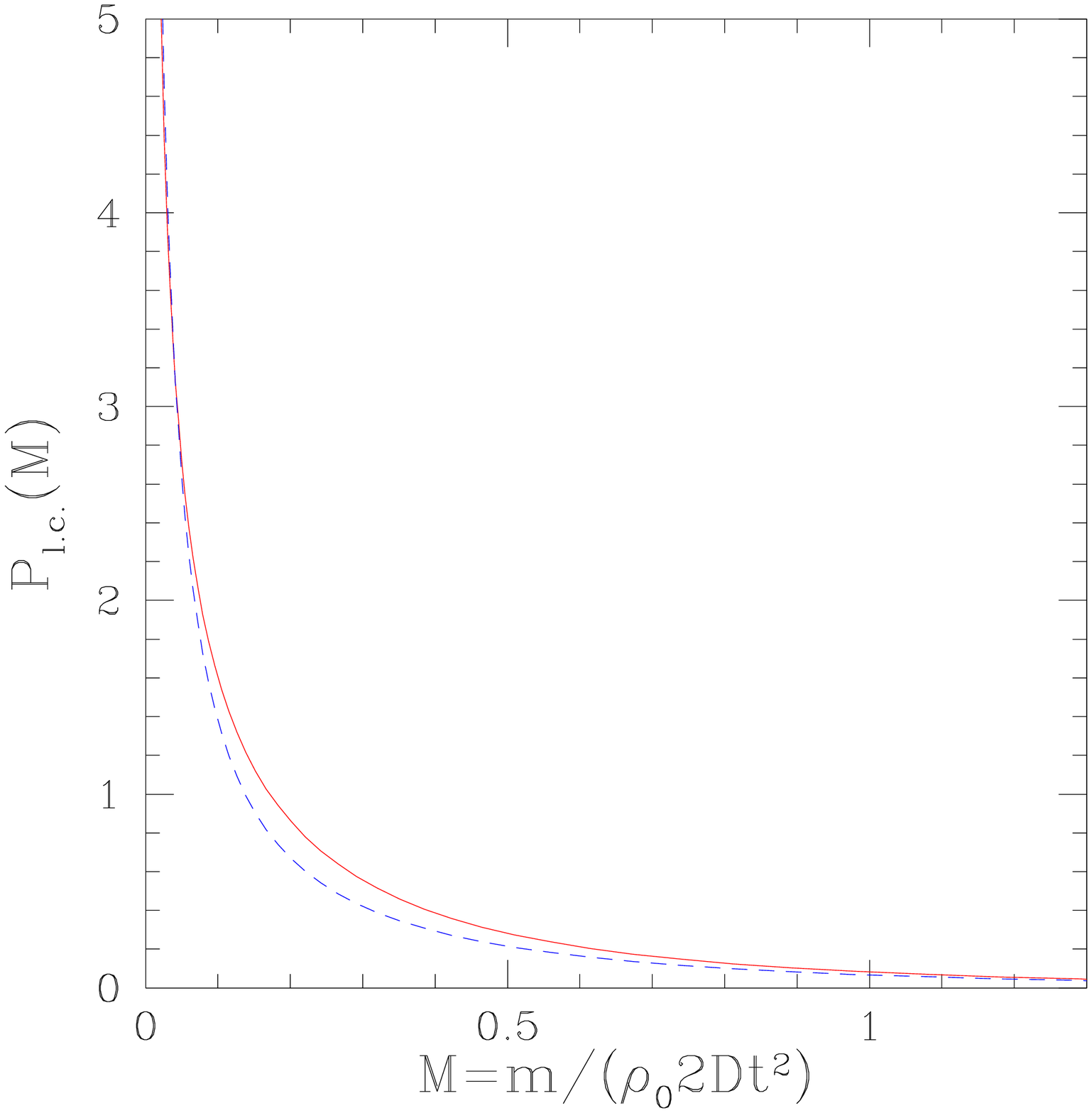}}
\epsfxsize=8 cm \epsfysize=6.2 cm {\epsfbox{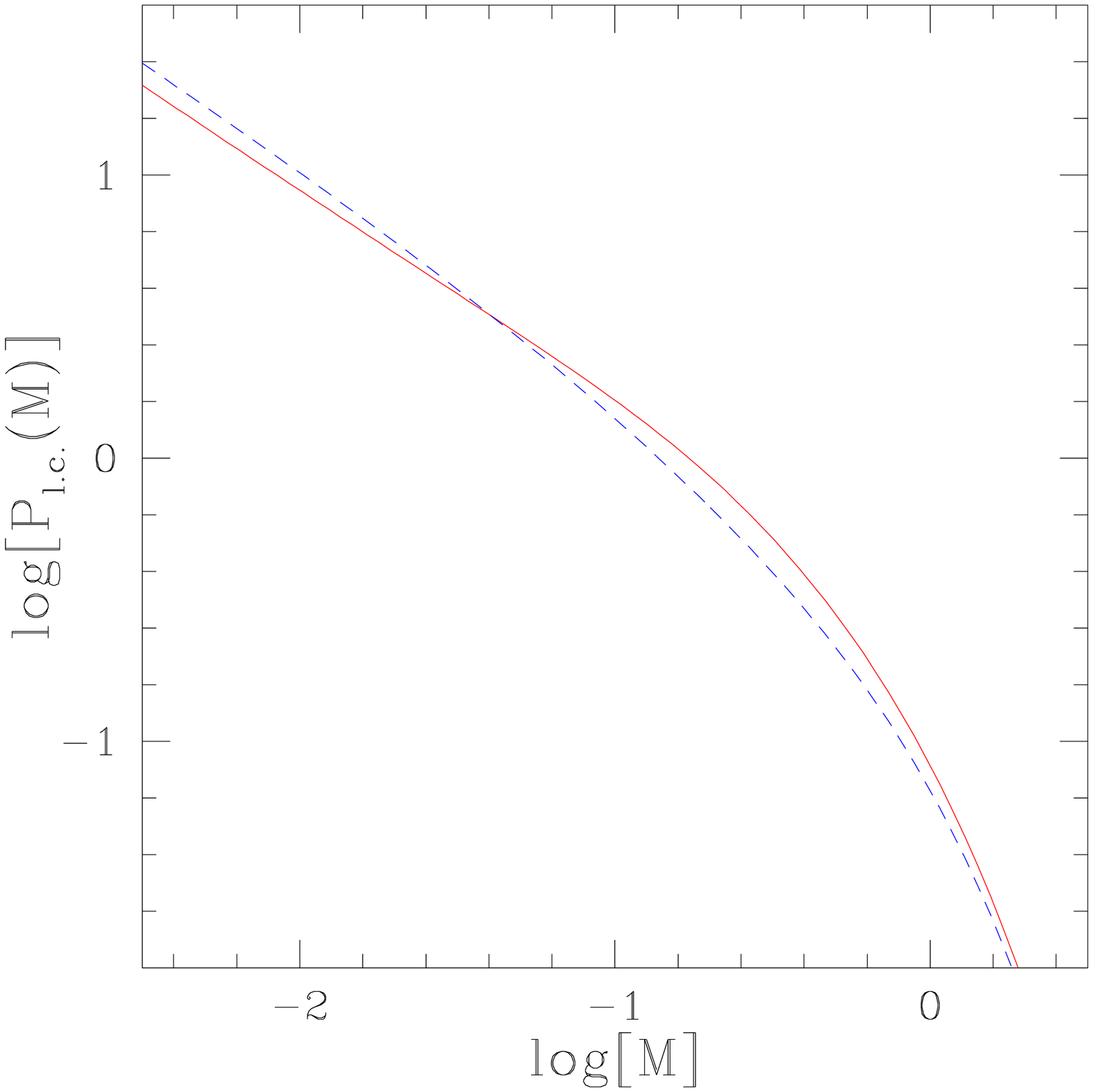}}
\end{center}
\caption{{\it Left panel:} The probability distribution $P_{\rm l.c.}(M)$ of the 
mass $M$ of the leftmost cluster.
We show our results for the filled case ``$F$'' (solid line, Eq.(\ref{PlcMF1}))
and the empty case ``$E$'' (dashed line, Eq.(\ref{PlcME1})).
{\it Right panel:} Same as left panel but on a logarithmic scale.}
\label{figPlcM}
\end{figure}

Next, to derive the distribution, $p_{\rm l.c.}(m)$, of the excited mass of this 
leftmost cluster, we first consider the bivariate distribution, 
$p_{\rm l.c.}(x,0\leq q'\leq q)\dd x$,
that this aggregate is located at a position in the range $[x,x+\dd x]$, with a right 
Lagrangian coordinate $q'$ that is smaller than $q$. This corresponds to a mass 
$m'=\rho_0 q'$ that is smaller than $m=\rho_0 q$. In a fashion similar to 
(\ref{pxcqF1}), we can write this quantity in terms of the Brownian propagators 
$K_{x,c}$ as
\beq
p_{\rm l.c.}(x,0\leq q'\leq q)\dd x = \lim_{q_+\rightarrow+\infty} 
\int \dd\psi\dd v \dd\psi_+\dd v_+
\, [ K_{x,0}(0,0,0;q,\psi,v) - K_{x+\dd x,0}(0,0,0;q,\psi,v) ] 
\, K_{x,0}(q,\psi,v;q_+,\psi_+,v_+) .
\label{plcxqF1}
\eeq
This equation states that the initial velocity potential $\psi_0$ obeys the following
two constraints, i) it stays above $\cP_{x,0}$ but goes below $\cP_{x+\dd x,0}$, over
the range $0\leq q'\leq q$, and ii) it remains above $\cP_{x,0}$ for $q'>q$.
Using the expressions given in Appendix A, as well as the results of appendices A 
and B of \cite{Valageas2008}, we obtain from Eq.(\ref{plcxqF1}), in terms of the 
reduced variables $X$ and $Q$,
\beqa
P_{\rm l.c.}(X,0\leq Q'\leq Q) & = & 2 e^{2X} \inta \frac{\dd s}{2\pi\ii} \, 
\frac{e^{(s-1)Q}}{s-1} \int_0^{\infty} \frac{\dd\nu}{\sqrt{\pi}} 3 \nu^{-3/2} 
e^{-\frac{2}{3}s^{3/2}\nu^{-3}-\nu^3X^2} \nonumber \\
&& \hspace{0cm} \times \biggl\lbrace \left(\sqrt{s}-\nu^3X\right)
\Ai\left[-\nu 2X+\frac{s}{\nu^2}\right] - \nu 
\Ai\,'\left[-\nu 2X+\frac{s}{\nu^2}\right] \biggl\rbrace , 
\;\;\; \mbox{with} \;\;\; \Re(s) > 1 .
\label{PlcXQF1}
\eeqa
We can check that the limit $Q\rightarrow+\infty$, which is governed by the pole at 
$s=1$ in Eq.(\ref{PlcXQF1}), yields back the probability density $P_{\rm l.c.}(X)$ 
given in Eq.(\ref{PlcXF}). Integrating over $X$, using an integration by parts and 
differentiating with respect to $Q$ gives the mass distribution
\beq
P_{\rm l.c.}(M) = \frac{M^{-3/4}}{\Gamma[1/4]} \, e^{-M} + \inta \frac{\dd s}
{2\pi\ii} \, e^{(s-1)M} (\sqrt{s}-1) \int_{-\infty}^0 \dd X \, 2 e^{2X} 
\int_0^{\infty} \frac{\dd\nu}{\sqrt{\pi}} 3 \nu^{-3/2} 
e^{-\frac{2}{3}s^{3/2}\nu^{-3}-\nu^3X^2} \Ai\left[-\nu 2X+\frac{s}{\nu^2}\right] ,
\label{PlcMF1}
\eeq
where the reduced mass is given by $M=m/(\rho_0\gam^2)=Q$, as in (\ref{Mdef}).
This yields the asymptotic behaviors
\beq
M \rightarrow 0^+ : \;\; P_{\rm l.c.}(M) \sim \frac{1}{\Gamma[1/4]} \, M^{-3/4} , 
\;\;\;\; M \rightarrow +\infty : \;\; P_{\rm l.c.}(M) \sim \alpha M^{-3/2} e^{-M}
\label{PlcMasympF1}
\eeq
with $\alpha= [ 2^{10/3} 3^{-13/3} \, _2F_1(\frac{5}{3},\frac{7}{6};\frac{5}{2};
\frac{5}{9}) + 2^{23/3} 3^{-17/3} \, _2F_1(\frac{7}{3},\frac{5}{6};\frac{5}{2};
\frac{5}{9}) ]/\sqrt{\pi} \simeq 0.511$.
We show in Fig.~\ref{figPlcM} our result (\ref{PlcMF1}) for the probability 
distribution of the mass of the leftmost cluster.
We can note that the high-mass tail has the same form, $M^{-3/2} e^{-M}$, as the
mass function of shocks on the right side $x>0$ (but with a slightly smaller
normalization), see \cite{Bertoin1998,Valageas2008}.
From Eq.(\ref{PlcMF1}), integrating over $M$ and $s$, we recognize the integral
in Eq.(\ref{XlcF}) and the mean excited mass of the leftmost cluster reads as
\beq
\lag m_{\rm l.c.}(t)\rag = \lag M_{\rm l.c.}\rag \rho_0 2D t^2 , \;\;\;
\mbox{with} \;\; \lag M_{\rm l.c.}\rag = \frac{1}{4} + \lag X_{\rm l.c.}\rag 
\simeq 0.19 .
\label{MlcF}
\eeq
Thus, the mass of the leftmost cluster grows with time as $t^2$ in the mean,
as it keeps being overtaken by new particles that arise from more distant regions
($q \sim t^2$) on the right side. We can note that the particles that were initially
at rest on the left side, and slow down its propagation, do not change the scaling 
laws associated with this cluster, as shown by Eq.(\ref{XlcF}). Indeed, since 
$x_{\rm l.c.} \sim -t^2$, the leftmost cluster has captured a mass 
$m^{\rm shocked} \sim t^2$ of still particles that were initially at rest on the
left side. On the other hand, it has just been overtaken by a mass $m \sim t^2$
of excited particles with $q\sim t^2$ and $v \sim -t$. Therefore, since the mass
$m^{\rm shocked}$ scales as $m$, the total momentum of the leftmost cluster
still scales as the momentum of the excited particles. As a consequence, we shall 
obtain in section~\ref{subsec/E:Leftmost_cluster} below the same scalings for the 
case ``$E$'' where the Brownian particles expand into empty space.

\section{Case ``$E$'': expansion into an empty medium}
\label{sec:case/E}

We now consider the case ``$E$'', defined in (\ref{Edef}), where the 
left side, $q<0$, is initially empty. We also compare our results with the
previous case ``$F$''.

\subsection{Eulerian velocity distribution and shock probability}
\label{subsec/E:Eulerian_velocity_distribution}

From the fact that the initial velocity potential is infinite on the left side,
$\psi_0(q)=+\infty$ for $q<0$, we can see that the computation 
(\ref{pxcqF1})-(\ref{PXpF}), performed in 
section~\ref{subsec:Eulerian_velocity_distribution} for the case of a filled left
side at rest, remains valid in the present case. Moreover, contrary to the previous
case, this same computation also holds on the left side, $x<0$, since the highest
height $c_*$ of the parabolas $\cP_{x,c}$ is still set by the origin 
$\{q=0,\psi_0=0\}$. Therefore, for all $X$ we obtain the probability distribution
of the Lagrangian coordinate $Q$ as:
\beq
\mbox{for any} \;\; X : \;\;  P_X(Q) = e^{2X} \inta \frac{\dd s}{2\pi\ii} 
\, e^{(s-1)Q} \int_0^{\infty} \frac{\dd\nu}{\sqrt{\pi}} 3 \nu^{-3/2} 
e^{-\frac{2}{3}s^{3/2}\nu^{-3}} \Ai\left[-\nu 2X +\frac{s}{\nu^2}\right] ,
\;\; \mbox{over} \;\; Q\geq 0 .
\label{PXE} 
\eeq
Of course, the Lagrangian coordinate $Q$ is always positive, since the left side
is initially empty.
Note that the only difference with Eq.(\ref{PXmF}) is the absence of the factor
$e^{-\nu^3 X^2}$, that arose from the additional constraint $\cP_{x,c}(q)\leq 0$
over $q<0$ in the filled left-side case ``$F$'' for $x<0$. 
For $X\geq 0$, we can integrate 
over $\nu$, using the results of Appendix~B in \cite{Valageas2008}, which gives 
back Eq.(\ref{PXpF}). Thus, all properties of the system on the right side, 
$X\geq 0$, are identical to those obtained in section~\ref{sec:case/F} for case 
``$F$''. Moreover, at large $Q$ and $V$, we recover the same asymptotic behaviors
(\ref{PXQVinfF}).

The shock probability, $P_X^{\rm shocked}$, that excited particles have already 
reached the position $X<0$ on the left side, now reads as
\beq
X \leq 0 : \;\; P_X^{\rm shocked} = e^{2X} \int_0^{\infty} 
\frac{\dd\nu}{\sqrt{\pi}} 3 \nu^{-3/2} e^{-\frac{2}{3}\nu^{-3}} 
\Ai\left[-\nu 2X +\frac{1}{\nu^2}\right] .
\label{PXshockedE} 
\eeq
Again, it only differs from Eq.(\ref{PXshockedF}) by the absence of the factor
$e^{-\nu^3 X^2}$. Since the Airy function $\Ai(y)$ is positive over $y\geq 0$
this shows that $P_X^{\rm shocked}$, as given by Eq.(\ref{PXshockedE}), is always
larger than the value (\ref{PXshockedF}) obtained in 
section~\ref{subsec:Eulerian_velocity_distribution} for the case ``$F$''.
This is due to the fact that in that previous case ``$F$'', particles that escape
from the right side are slowed down as they travel to the left by the matter that
was initially at rest on the left side. (More precisely, this matter slows down the
leftmost cluster, which also slows down particles that overtake it and aggregate to
it.) This clearly implies that, at any location $X<0$, $P_X^{\rm shocked}$ is smaller
for the filled left-side case ``$F$'', as can be checked in Fig.~\ref{figPshock}
where we compare the result (\ref{PXshockedE}) (dashed line) with the previous
result (\ref{PXshockedF}) (solid line).
From Eq.(\ref{PXshockedE}) we obtain the asymptotic behaviors
\beq
X \rightarrow 0^- : \;\; P_X^{\rm shocked} \sim 1 - \sqrt{\frac{-24X}{\pi}} , 
\;\;\;\;\;\; X \rightarrow -\infty : \;\; P_X^{\rm shocked} \sim 
\frac{1}{\sqrt{-6\pi X}} \,\, e^{6X} .
\label{PXshockedasympE}
\eeq
We can check that they are larger than the results (\ref{PXshockedasympF})
obtained for case ``$F$'', in agreement with the discussion above.
We can note that the exponent $1/3$ has been changed to $1/2$ for the limit
$X\rightarrow 0^-$, whereas the large-$X$ decay still has the form of an exponential
multiplied by an inverse square-root, but with a different numerical factor in the
exponential.

\subsection{Mean density profile and mean current}
\label{subsec/E:density-profile}

As discussed in section~\ref{subsec/E:Eulerian_velocity_distribution},
the properties of the system, whence the mean density and current, are identical
over $x\geq 0$ to those obtained for the previous filled left-side case in 
section~\ref{sec:case/F}. In particular, we recover the results 
(\ref{meanQVpF})-(\ref{meanMrhojpF}). Therefore, we now focus on the left side 
$x<0$. For the present case, where there are no particles at $x<0$ until some 
particles from the right side have travelled down to position $x$, the mean 
Lagrangian coordinate $\lag q(x)\rag$ and the mean velocity $\lag v(x)\rag$ are not
meaningful. However, we can still define the mean mass $\lag m(<x)\rag$ of excited
particles located to the left of the Eulerian position $x$, as well as the mean
density and mean current (all these quantities being equal to zero until particles
from the right semi-infinite axis have reached position $x$). Then, we obtain 
from Eq.(\ref{PXE})
\beq
X\leq 0 : \;\; \lag M(<X)\rag= e^{2X} \int_0^{\infty} \frac{\dd\nu}{\sqrt{\pi}} 
3 \nu^{-3/2} e^{-\frac{2}{3}\nu^{-3}} \left[ \nu^{-3} \, 
\Ai\left(-\nu 2X +\nu^{-2}\right) - \nu^{-2}\, \Ai\,'\left(-\nu 2X +\nu^{-2}\right) 
\right]  ,
\label{meanMmE}
\eeq
\beq
x\leq 0 : \;\; \lag \rho(x)\rag= \rho_0 \, e^{2X} \int_0^{\infty} 
\frac{\dd\nu}{\sqrt{\pi}} 3 \nu^{-3/2} e^{-\frac{2}{3}\nu^{-3}} 
\left[ (4\nu^{-3}-4X) \, \Ai\left(-\nu 2X +\nu^{-2}\right) - 4\nu^{-2} \, 
\Ai\,'\left(-\nu 2X +\nu^{-2}\right) \right] ,
\label{meanrhomE}
\eeq
and again for $x\leq 0$,
\beq
\lag j(x)\rag= -4\rho_0 D t \, e^{2X} \int_0^{\infty} 
\frac{\dd\nu}{\sqrt{\pi}} 3 \nu^{-3/2} e^{-\frac{2}{3}\nu^{-3}} 
\left[ (\nu^{-3}-4\nu^{-3}X+4X^2) \, \Ai\left(-\nu 2X +\nu^{-2}\right) 
- (\nu^{-2}-4\nu^{-2}X) \, \Ai\,'\left(-\nu 2X +\nu^{-2}\right) \right] .
\label{meanjmE}
\eeq
This yields close to the boundary $x=0$ the asymptotic behaviors 
\beqa
X \rightarrow 0^- & : & \lag M(<X)\rag \sim \frac{1}{4} + X +
 \frac{2^{9/2}3^{1/2}}{5\sqrt{\pi}} (-X)^{5/2} , \;\;\;\; \lag \rho(x)\rag \sim 
\rho_0 \left[ 1 - \frac{2^{7/2}3^{1/2}}{\sqrt{\pi}} (-X)^{3/2} \right] , 
\nonumber \\ 
&& \lag j(x)\rag \sim - \rho_0 D t \left[ 1 - \frac{2^{11/2}3^{3/2}}{5\sqrt{\pi}} 
(-X)^{5/2} \right] .
\label{meanMrhojX0asympmE}
\eeqa
Thus, we can see that the transitory increase obtained for the amplitude of the
mean density and current in the case of the filled left side, see 
Eq.(\ref{meanMrhojX0asympmF}), that was associated with the dragging effect due
to matter that was initially at rest on the left side,
is no longer present when the Brownian particles expand into an empty medium.
At large distance we obtain the same exponential decay as for the shock probability 
$P_X^{\rm shocked}$,
\beq
X \rightarrow -\infty : \;\; \lag M(<X)\rag \sim \sqrt{\frac{-2X}{3\pi}} \,
e^{6X} , \;\; \lag \rho(x)\rag \sim \rho_0 \, 2 \sqrt{\frac{-6X}{\pi}} \,
e^{6X} , \;\; \lag j(x)\rag \sim - \rho_0 D t \, 8 \sqrt{\frac{6}{\pi}} \, 
(-X)^{3/2} \, e^{6X} .
\label{meanMrhojasympmE}
\eeq
Again, we can check that, far from $x=0$, the amplitude of the mean mass, density 
and current are larger than for the filled left-side case given in 
(\ref{meanVMrhojasympmF}). This agrees with the larger value of $P_X^{\rm shocked}$
discussed above. We can check these properties in Figs.~\ref{figrho}, 
\ref{figjmean}, where we compare the mean density and current obtained for both 
cases.

We can note that the propagation into empty medium for the Brownian case studied
here shows significantly different properties from the case of white-noise initial
velocity studied in \cite{Frachebourg2001}. Indeed, in the latter case, the 
typical distances and masses only scale as $t^{2/3}$, in a fashion similar to the
scaling $t^2$ associated with the scaling laws (\ref{scalings}) that express
the scale invariance of the initial velocity field. However, the mean mass, 
$\lag M(<X)\rag$, located to the left of $X$, only decays as the inverse 
power-law $1/X^2$ (with now $X\sim x/t^{2/3}$ and $M \sim m/t^{2/3}$), instead
of the exponential falloff (\ref{meanMrhojasympmE}). Then, one still obtains a finite
mass distribution in the ballistic limit $t\rightarrow \infty$ for $p_{\xi t}(m,t)$ 
at fixed $\xi$ and $m$. This corresponds to free-moving forerunners that carry 
a finite mass in front of the typical profile that extends on the smaller scale
$\sim t^{2/3}$. For the case when particles spread into a filled medium at rest,
this distribution vanishes \cite{Frachebourg2001}. In the present Brownian case,
the appearance of such a new nontrivial scaling, specific to the expansion into
empty space, is no longer possible, as shown by the exponential decay 
(\ref{meanMrhojasympmE}) and the results derived above. Indeed, the front now 
shows a faster than ballistic propagation in both cases ``$F$'' and ``$E$'' --
as $t^2$ as given by the scalings (\ref{scalings}) -- which is governed by the latest
high-velocity particles coming from increasingly far regions on the right (see
the discussion below Eq.(\ref{MlcF})). Then, this super-ballistic propagation  
leaves no room for new scalings and the dynamics remains very similar for
both cases ``$F$'' and ``$E$'' (see also \cite{Isozaki2006} for a comparison
of white-noise and Brownian initial conditions).

\subsection{Leftmost cluster}
\label{subsec/E:Leftmost_cluster}

Within a Lagrangian point of view, the probability distributions of the displacement
field are again given by Eq.(\ref{PQXpF}) over $X\geq 0$, and by Eq.(\ref{PQXmF})
over $X\leq 0$ but without the factor $e^{-\nu^3 X^2}$, in agreement with the
previous discussion. Then, we focus here on the leftmost cluster, which is also
associated with the particle $q=0$. From Eq.(\ref{PXshockedE}) we now obtain
\beq
X < 0 : \;\; P_{\rm l.c.}(X) = \frac{\dd}{\dd X} P_X^{\rm shocked} = 
\frac{\dd}{\dd X} \, e^{2X} \int_0^{\infty} \frac{\dd\nu}{\sqrt{\pi}} 3 \nu^{-3/2} 
e^{-\frac{2}{3}\nu^{-3}} \Ai\left[-\nu 2X+\frac{1}{\nu^2}\right] ,
\label{PlcXE}
\eeq
which leads to the asymptotic behaviors
\beq
X \rightarrow 0^- : \;\; P_{\rm l.c.}(X) \sim \sqrt{\frac{6}{-\pi X}} , 
\;\;\;\; X \rightarrow -\infty : \;\; P_{\rm l.c.}(X) \sim \sqrt{\frac{6}{-\pi X}}
\, e^{6X} ,
\label{PlcXasympE}
\eeq
while the mean position is
\beq
\lag x_{\rm l.c.}(t)\rag = \lag X_{\rm l.c.}\rag 2D t^2 \;\;\;\; \mbox{with} 
\;\;\;\; \lag X_{\rm l.c.}\rag - \int_{-\infty}^0 \dd X \, e^{2X} \int_0^{\infty}
\frac{\dd\nu}{\sqrt{\pi}} 3 \nu^{-3/2} e^{-\frac{2}{3}\nu^{-3}} 
\Ai\left[-\nu 2X+\frac{1}{\nu^2}\right]  \simeq-0.08  .
\label{XlcE}
\eeq
As discussed in section~\ref{subsec:Leftmost_cluster}, we recover the same scaling
law as for the case where Brownian particles expand into a filled medium at rest,
see Eq.(\ref{XlcF}). However, the mean of the reduced variable $X_{\rm l.c.}$ has
a slightly larger absolute value, since it is easier for the leftmost cluster to 
travel to the far left as it is no longer slowed down by the particles that were 
initially at rest in the filled case ``$F$''. 
This also leads to the smaller low-$X$ tail and to the larger high-$X$ tail that 
can be seen in Fig.~\ref{figPlcX} and by the comparison of Eq.(\ref{PlcXasympE}) 
with Eq.(\ref{PlcXasympF}). Note however that for the most part both distributions
$P_{\rm l.c.}(X)$ are very close to each other.

The bivariate distribution $p_{\rm l.c.}(x,0\leq q'\leq q)\dd x$ is now given by
\beqa
p_{\rm l.c.}(x,0\leq q'\leq q)\dd x & = & \lim_{q_+\rightarrow+\infty} 
\int \dd\psi\dd v \dd\psi_+\dd v_+
\, [K_{x,x^2/(2t)}(0,0,0;q,\psi,v) - K_{x+\dd x,x^2/(2t)+(x/t)\dd x}(0,0,0;q,\psi,v)] 
\nonumber \\
&& \times \, K_{x,x^2/(2t)}(q,\psi,v;q_+,\psi_+,v_+) .
\label{plcxqE1}
\eeqa
Indeed, since $\psi_0(q)=+\infty$ over $q<0$, we no longer have to consider parabolas
$\cP_{x,c}$ with $c=0$, which are tangent to the horizontal axis $\psi_0=0$, as
we did in Eq.(\ref{plcxqF1}) for the case ``$F$''. In the present case ``$E$'', we
must consider parabolas which run through the origin, $\cP_{x,c}(0)=0$, 
whence $c=x^2/(2t)$. Using the expressions given in Appendix A, as well as the 
results of appendices A and B of \cite{Valageas2008}, we obtain from 
Eq.(\ref{plcxqE1})
\beq
P_{\rm l.c.}(X,0\leq Q'\leq Q) = 2 e^{2X} \inta \frac{\dd s}{2\pi\ii} \, 
\frac{e^{(s-1)Q}}{s-1} \int_0^{\infty} \frac{\dd\nu}{\sqrt{\pi}} 3 \nu^{-3/2} 
e^{-\frac{2}{3}s^{3/2}\nu^{-3}} \biggl\lbrace \sqrt{s}
\Ai\left[-\nu 2X+\frac{s}{\nu^2}\right] - \nu 
\Ai\,'\left[-\nu 2X+\frac{s}{\nu^2}\right] \biggl\rbrace ,
\label{PlcXQE1}
\eeq
where the integration contour obeys $\Re(s)>1$. Again, taking the limit 
$Q\rightarrow +\infty$ we can check that we recover the distribution (\ref{PlcXE}).
Then, integrating over $X$ yields the mass distribution
\beq
P_{\rm l.c.}(M) = \frac{M^{-3/4}}{\Gamma[1/4]} \, e^{-M} + \inta \frac{\dd s}
{2\pi\ii} \, e^{(s-1)M} (\sqrt{s}-1) \int_{-\infty}^0 \dd X \, 2 e^{2X} 
\int_0^{\infty} \frac{\dd\nu}{\sqrt{\pi}} 3 \nu^{-3/2} 
e^{-\frac{2}{3}s^{3/2}\nu^{-3}} \Ai\left[-\nu 2X+\frac{s}{\nu^2}\right] .
\label{PlcME1}
\eeq
This gives the asymptotic behaviors
\beq
M \rightarrow 0^+ : \;\; P_{\rm l.c.}(M) \sim \frac{\sqrt{3/2}}{\Gamma[1/4]} \,
M^{-3/4}, \;\;\;\; M \rightarrow +\infty : \;\; P_{\rm l.c.}(M) \sim 
\frac{\sqrt{3/\pi}}{2} \, M^{-3/2} e^{-M} ,
\label{PlcMasympE1}
\eeq
and the mean mass
\beq
\lag m_{\rm l.c.}(t)\rag = \lag M_{\rm l.c.}\rag \rho_0 2D t^2 , \;\;\;
\mbox{with} \;\; \lag M_{\rm l.c.}\rag = \frac{1}{4} + \lag X_{\rm l.c.}\rag 
\simeq 0.17 ,
\label{MlcE}
\eeq
where we recognize the integral (\ref{XlcE}), in a fashion similar to
(\ref{MlcF}).
Thus, as for the position $x_{\rm l.c.}(t)$, we recover the same scalings for the
mass $m_{\rm l.c.}(t)$ as for the case ``$F$'' of the expansion into a filled medium
at rest. The mean reduced mass $\lag M_{\rm l.c.}\rag$ is slightly smaller than for
the case ``$F$''. Indeed, since the leftmost cluster is no longer slowed down by
particles that were initially at rest on the left side, it moves somewhat farther
into the left side, as seen in Eq.(\ref{XlcE}), which implies that fewer particles
from the right side have been able to overtake it. This leads to a smaller mass
of excited particles that have been able to aggregate into this cluster.
This now implies a larger low-$M$ tail
and a smaller high-$M$ tail, as compared with the case ``$F$''.  
These properties can be checked in Fig.~\ref{figPlcM} and by the comparison of
Eq.(\ref{PlcMasympE1}) with Eq.(\ref{PlcMasympF1}). However, as for the position
$X_{\rm l.c.}$, both distributions $P_{\rm l.c.}(M)$ remain for the most part
very close to each other.

\section{Conclusion}
\label{sec:Conclusion}

In this article we have studied the one-dimensional ballistic aggregation process,
in the continuum limit and for the case where the initial velocity on the right 
semi-infinite axis is a Brownian motion. The left side is either at rest, with the
same uniform initial density, or empty. Then, focussing on the out-of-equilibrium
propagation of particles towards the left of the system, we noticed that in both
cases the mean density remains constant on the right side whereas a mean current
towards the left develops and grows linearly with time. Thus, particles coming
from increasingly far regions on the right replenish the system, as seen at any 
finite distance $x>0$ to the right of the origin, and balance the mean loss of 
matter associated with particles that have escaped into the left semi-infinite 
axis. Moreover, the properties of velocity increments, of the density field and of
shocks are the same in both cases and are also identical to those obtained 
asymptotically far from the origin for the case of two-sided Brownian initial 
conditions.

We find that on the far left, for both cases all quantities (e.g. mean 
density and current) show an exponential decay, with a slightly faster falloff 
for the case ``$F$'' where the particles expand into a filled medium at rest. 
Indeed, in this latter case, the initially still particles that were located on 
the left side slow down the propagation towards the left of the leftmost cluster, 
built by the boundary particle, $q=0$, and by all particles $q>0$ that have already 
overtaken it, as they aggregate to it and decrease its total momentum.
This dragging effect also leads to a transitory increase to the left of the
boundary $x=0$ of the mean density and current, which is not present when particles
expand into empty space. This also leads to a leftmost cluster which is statistically
closer to the boundary $x=0$ and more massive (counting only the particles that
came from the right part) as it is easier for particles from the right side to
overtake it. 

An interesting feature of this one-dimensional process is that it provides a 
nontrivial inhomogeneous non-equilibrium system where many quantities can
be obtained explicitly, as seen for instance in the calculations presented here.
The same methods could also be applied to different time statistics, but we leave
such studies for future works. Then, the simple system described here may also be 
used as a benchmark to test approximation schemes devised for more difficult cases
where it is not possible to derive exact results.

\appendix

\section{Brownian propagators}
\label{app:Brownian_propagators}

The reduced propagator $G$ introduced in Eq.(\ref{KG1}) is most easily written in
terms of its Laplace transform, $\tG$, defined by \cite{Valageas2008}
\beq
\tG(s;r_1,u_1;r_2,u_2) = \int_0^{\infty} \dd\tau \, e^{-s\tau} 
G(\tau;r_1,u_1;r_2,u_2) , \;\;\; \mbox{and} \;\;\; \tG=\tG_0-\tG_1 \;\; \mbox{with}
\label{tGdef}
\eeq
\beq
\tG_0(s;r_1,u_1;r_2,u_2) = \int_{-\infty}^{\infty} \dd\nu \, 
e^{-\nu^3(r_2-r_1)} \, 3\nu \, \Ai\left[-\nu u_1 +\frac{s}{\nu^2}\right] 
\Ai\left[-\nu u_2 +\frac{s}{\nu^2}\right] \left[ - \theta(-\nu) \theta(r_1-r_2)
+ \theta(\nu) \theta(r_2-r_1) \right] ,
\label{tG0}
\eeq
\beq
\tG_1 = \int_0^{\infty} \frac{\dd\nu \dd\mu}{2\pi} \, 
\frac{9\nu^{3/2}\mu^{3/2}}{\nu^3+\mu^3} 
\, e^{-\frac{2}{3} s^{3/2} (\nu^{-3}+\mu^{-3})} \, e^{-\nu^3 r_1-\mu^3 r_2} 
\, \Ai\left[\nu u_1+\frac{s}{\nu^2}\right] 
\Ai\left[-\mu u_2+\frac{s}{\mu^2}\right] .
\label{tG1}
\eeq
The term $G_0$ actually corresponds to unconstrained Brownian trajectories,
hence it is also given by \cite{Burkhardt1993}
\beq
G_0(\tau;r_1,u_1;r_2,u_2) = \frac{\sqrt{3}}{2\pi\tau^2} \, 
e^{-\frac{3}{\tau^3}(r_2-r_1-u_1\tau)^2
+\frac{3}{\tau^2}(r_2-r_1-u_1\tau)(u_2-u_1)-\frac{1}{\tau}(u_2-u_1)^2} .
\label{G0tau}
\eeq
From the propagator $G$ it is convenient to derive the kernel $\Hi$,
associated with Brownian particles that remain forever above the parabola 
$\cP_{x,c}$,
\beq
\lim_{q_2\rightarrow+\infty} \int \dd\psi_2\dd v_2 \, 
K_{x,c}(q_1,\psi_1,v_1;q_2,\psi_2,v_2) = e^{-u_1/\gamma} \Hi(r_1,u_1) .
\label{KHi}
\eeq
We also consider the propagators, $\Delta$ and $E$, associated with Brownian 
particles that come within a small vertical distance $\delta c$, or horizontal 
distance $\delta x$, from the parabolic absorbing barrier:
\beq
\lim_{\delta c \rightarrow 0} \, \frac{1}{\delta c}
[K_{x,c}(q_1,\psi_1,v_1;q_2,\psi_2,v_2) 
- K_{x,c+\delta c}(q_1,\psi_1,v_1;q_2,\psi_2,v_2)] \, \dd\psi_2\dd v_2 = 
2\frac{t}{\gamma} \, e^{-\tau/\gamma^2+(u_2-u_1)/\gamma} 
\, \Delta(\tau;r_1,u_1;r_2,u_2) \, \dd r_2 \dd u_2 ,
\label{KDelta}
\eeq
and
\beqa
\lim_{\delta x \rightarrow 0} \, \frac{1}{\delta x}
[K_{x,c}(q_1,\psi_1,v_1;q_2,\psi_2,v_2) 
- K_{x+\delta x,c}(q_1,\psi_1,v_1;q_2,\psi_2,v_2)] \, \dd\psi_2\dd v_2 & = & 
\nonumber \\
&& \hspace{-5cm} 2\gamma^{-1} \, e^{-\tau/\gamma^2+(u_2-u_1)/\gamma} 
\, E(\tau;r_1,u_1;r_2,u_2;q_1,q_2,x) \, \dd r_2 \dd u_2 .
\label{KE}
\eeqa
Using Eqs.(\ref{tG0})-(\ref{tG1}), one obtains the expressions
\beq
\Hi(r_1,u_1) = e^{u_1/\gam} - \int_0^{\infty} \frac{\dd\nu}{\sqrt{\pi}} 
3 \nu^{-3/2} e^{-\frac{2}{3}\nu^{-3}-\nu^3 r_1/\gam^3} 
\Ai\left[\nu \frac{u_1}{\gam}+\frac{1}{\nu^2}\right] ,
\label{Hieq}
\eeq
\beq
\tD(s;r_1,u_1;r_2,u_2) = \int_0^{\infty} \frac{\dd\nu\dd\mu}{2\pi} \, 
9\nu^{3/2}\mu^{3/2} \, e^{-\frac{2}{3} s^{3/2} (\nu^{-3}+\mu^{-3})} 
\, e^{-\nu^3 r_1-\mu^3 r_2}  
\Ai\left[\nu u_1+\frac{s}{\nu^2}\right] 
\Ai\left[-\mu u_2+\frac{s}{\mu^2}\right]  ,
\label{Ds}
\eeq
and
\beqa
\tE(s;r_1,u_1;r_2,u_2;q_1,q_2,x) & = & - \int_0^{\infty} \frac{\dd\nu\dd\mu}{2\pi} \, 
\frac{9\nu^{3/2}\mu^{3/2}}{\nu^3+\mu^3} \, 
e^{-\frac{2}{3} s^{3/2} (\nu^{-3}+\mu^{-3})} \, e^{-\nu^3 r_1-\mu^3 r_2} 
\biggl\lbrace \nu \, \Ai\,'\left[\nu u_1+\frac{s}{\nu^2}\right]
\Ai\left[-\mu u_2+\frac{s}{\mu^2}\right] \nonumber \\
&& \hspace{-3cm} - \mu \, \Ai\left[\nu u_1+\frac{s}{\nu^2}\right] 
\Ai\,'\left[-\mu u_2+\frac{s}{\mu^2}\right] + \left[\nu^3(x-q_1)+\mu^3(x-q_2)\right] 
\, \Ai\left[\nu u_1+\frac{s}{\nu^2}\right] \Ai\left[-\mu u_2+\frac{s}{\mu^2}\right] 
\biggl \rbrace .
\label{Es}
\eeqa


\begin{thebibliography}{99}

\bibitem[Aurell et al. (1993)]{Aurell1993}
E. Aurell, S. N. Gurbatov, I. I. Wertgeim, 1993, Phys. Lett. A, 182, 109

\bibitem[Bec \& Khanin (2007)]{Bec2007}
J. Bec, K. Khanin, 2007, Phys. Rep., 447, 1

\bibitem[Bertoin (1998)]{Bertoin1998}
J. Bertoin, 1998, Commun. Math. Phys., 193, 397

\bibitem[Bertoin (2000)]{Bertoin2000}
J. Bertoin, 2000, J. Math. Pures Appl., 79, 173

\bibitem[Bertoin et al. (2001)]{Bertoin2001}
J. Bertoin, C. Giraud, Y. Isozaki, 2001, Commun. Math. Phys., 224, 551

\bibitem[Burgers (1974)]{Burgersbook}
J. M. Burgers, 1974, The nonlinear diffusion equation, D. Reidel, Dordrecht

\bibitem[Burkhardt (1993)]{Burkhardt1993}
T. W. Burkhardt, 1993, J. Phys. A, 26, L1157

\bibitem[Carnevale et al. (1990)]{Carnevale1990}
G. F. Carnevale, Y. Pomeau, W. R. Young, 1990, Phys. Rev. Lett., 64, 2913

\bibitem[Cole (1951)]{Cole1951}
J. D. Cole, 1951, Quart. Appl. Math., 9, 225

\bibitem[Frachebourg (1999)]{Frachebourg1999}
L. Frachebourg, 1999, Phys. Rev. Lett., 82, 1502

\bibitem[Frachebourg \& Martin (2000)]{Frachebourg2000}
L. Frachebourg, Ph. A. Martin, 2000, J. Fluid Mech., 417, 323

\bibitem[Frachebourg et al. (2000)]{Frachebourg2000a}
L. Frachebourg, Ph. A. Martin, J. Piasecki, 2000, Physica A, 279, 69

\bibitem[Frachebourg et al. (2001)]{Frachebourg2001}
L. Frachebourg, V. Jacquemet, Ph. A. Martin, 2001, J. Stat. Phys., 105, 745

\bibitem[Gurbatov et al. (1989)]{Gurbatov1989}
S. N. Gurbatov, A. I. Saichev, S. F. Shandarin, 1989, Mont. Not. Roy. Astron. Soc., 236, 385

\bibitem[Gurbatov et al. (1991)]{Gurbatov1991}
S. Gurbatov, A. Malakhov, A. Saichev, 1991, Nonlinear random waves and turbulence in nondispersive media: waves, rays and particles, Manchester University Press

\bibitem[Gurbatov et al. (1997)]{Gurbatov1997}
S. N. Gurbatov, S. I. Simdyankin, E. Aurell, U. Frisch, G. Toth, 1997, J. Fluid Mech., 344, 339

\bibitem[Gurbatov \& Pasmanik (1999)]{Gurbatov1999}
S. N. Gurbatov, G. V. Pasmanik, 1999, Sov. Phys. JETP, 88, 309

\bibitem[Hopf (1950)]{Hopf1950}
E. Hopf, 1950, Commun. Pure Appl. Mech., 3, 201

\bibitem[Isozaki (2006)]{Isozaki2006}
Y. Isozaki, 2006, Osaka J. Math., 43, 239 

\bibitem[Kida (1979)]{Kida1979}
S. Kida, 1979, J. Fluid Mech., 93, 337

\bibitem[Majumdar et al. (2008)]{Majumdar2008}
S. Majumdar, K. Mallick, S. Sabhapandit, 2008, arXiv:0911.0908

\bibitem[Molchan (1997)]{Molchan1997}
G. M. Molchan, 1997, J. Stat. Phys., 88, 1139

\bibitem[Molchanov et al. (1995)]{Molchanov1995}
S. A. Molchanov, D. Surgailis, W.A. Woyczynski, 1995, Commun. Math. Phys., 168, 209

\bibitem[She et al. (1992)]{She1992}
Z.-S. She, E. Aurell, U. Frisch, 1992, Commun. Math. Phys., 148, 623

\bibitem[Sinai (1992)]{Sinai1992}
Ya. G. Sinai, 1992, Commun. Math. Phys., 148, 601

\bibitem[Suidan (2000)]{Suidan2000}
T. Suidan, 2000, J. Stat. Phys., 101, 893

\bibitem[Valageas (2008)]{Valageas2008}
P. Valageas, 2008, accepted by J. Stat. Phys., arXiv:0810.4332

\bibitem[Vergassola et al. (1994)]{Vergassola1994}
M. Vergassola, B. Dubrulle, U. Frisch, A. Noullez, 1994, Astron. Astrophys., 289, 325

\bibitem[Winkel (2002)]{Winkel2002}
M. Winkel, 2002, J. Stat. Phys., 107, 893



\end{thebibliography}
\end{document}